\documentclass[11pt]{article}
\usepackage[utf8]{inputenc}
\usepackage{graphicx}
\usepackage{natbib}
\RequirePackage{amsthm,amsmath,amsfonts,amssymb}
\usepackage[dvipsnames]{xcolor}
\usepackage{bbm}
\usepackage[font={small},labelfont=it,textfont=it]{caption}
\usepackage[bottom]{footmisc}
\usepackage{tikz}
\usetikzlibrary{arrows.meta,positioning,calc,decorations.pathreplacing,shapes.geometric}
\usepackage{mathtools}
\usepackage{enumerate}
\usepackage[margin=1in]{geometry}
\usepackage{multicol}
\usepackage{booktabs}
\usepackage{url}
\usepackage{hyperref}
\usepackage{subcaption}
\usepackage{chngcntr}
\usepackage{floatrow}
\DeclareFloatSeparators{mysep}{\hskip-6em}

\hypersetup{pdfborder = {0 0 0.5 [3 3]}, colorlinks = true, linkcolor = BrickRed, citecolor = SkyBlue}

\DeclareMathOperator*{\argmax}{arg\,max}

\numberwithin{equation}{section}

\newcommand{\appref}[2]{\ref{#1}}
\newcommand{\insupp}{}
\newcommand{\mainref}[2]{\ref{#1}}
\newcommand{\appsec}[2]{Appendix~\ref{#1}}

\newcommand{\authcell}[2]{%
  \begin{minipage}[t]{0.3\textwidth}\centering
    \textbf{#1}\\[0.2ex]{\footnotesize\linespread{0.9}\selectfont #2\par}%
  \end{minipage}}

\title{Stated, Realized, and Optimal Aiming Strategies for the Tennis Serve: Experimental Evidence from Collegiate Athletes}

\author{%
\authcell{Nathan Sandholtz\thanks{Contact email: nsandholtz@stat.byu.edu}}{Department of Statistics,\\ Brigham Young University}\hspace{1.5em}%
\authcell{Ryan Hanson}{}\hspace{1.5em}%
\authcell{Ronald L. Hager}{Department of Exercise Sciences,\\ Brigham Young University}\\[7ex]
\authcell{Stephanie Kovalchik}{Teamworks Innovations, Inc.}\hspace{1.5em}%
\authcell{Gilbert W. Fellingham}{Department of Statistics,\\ Brigham Young University}%
}

\date{}

\begin{document}

\maketitle

\begin{abstract}

We study how a player’s execution error affects the optimal location at which to aim a serve in tennis. Aiming near the service-box boundaries makes a serve harder to return but increases the probability of a fault, creating a trade-off that depends on the player's execution error. Match-play data provide limited information about execution error because players’ intended targets are unobserved. We conducted an experiment with Brigham Young University men’s tennis players in which players explicitly identified targets they believed to be optimal for their serves. Players then served repeatedly toward their chosen targets, and we recorded the bounce locations.

Using these data, we estimate player-specific execution error via a Bayesian hierarchical model that accounts for serves censored by net contact. These distributions define the transition dynamics of a two-period Markov decision process (MDP) representing a point in tennis, where the decision variables are the first- and second-serve aim locations. We solve the MDP for each posterior draw of the execution error distributions, yielding a posterior distribution over each player’s optimal aim location. We find that players’ stated targets are generally more aggressive than optimal, whereas their realized serve centers are more conservative and markedly closer to the optima.

 \end{abstract}

\providecommand{\studyteam}{Brigham Young University Men's Tennis Team}
\providecommand{\studyirb}{Brigham Young University Institutional Review Board (IRB\#~2025-487)}
\providecommand{\anon}{1}

\section{Introduction}
\label{sec:introduction}

\begin{figure}[h]
    \centering
    \includegraphics[trim = 1.25in 1.25in .5in 1.9in, clip, width=.98\textwidth]{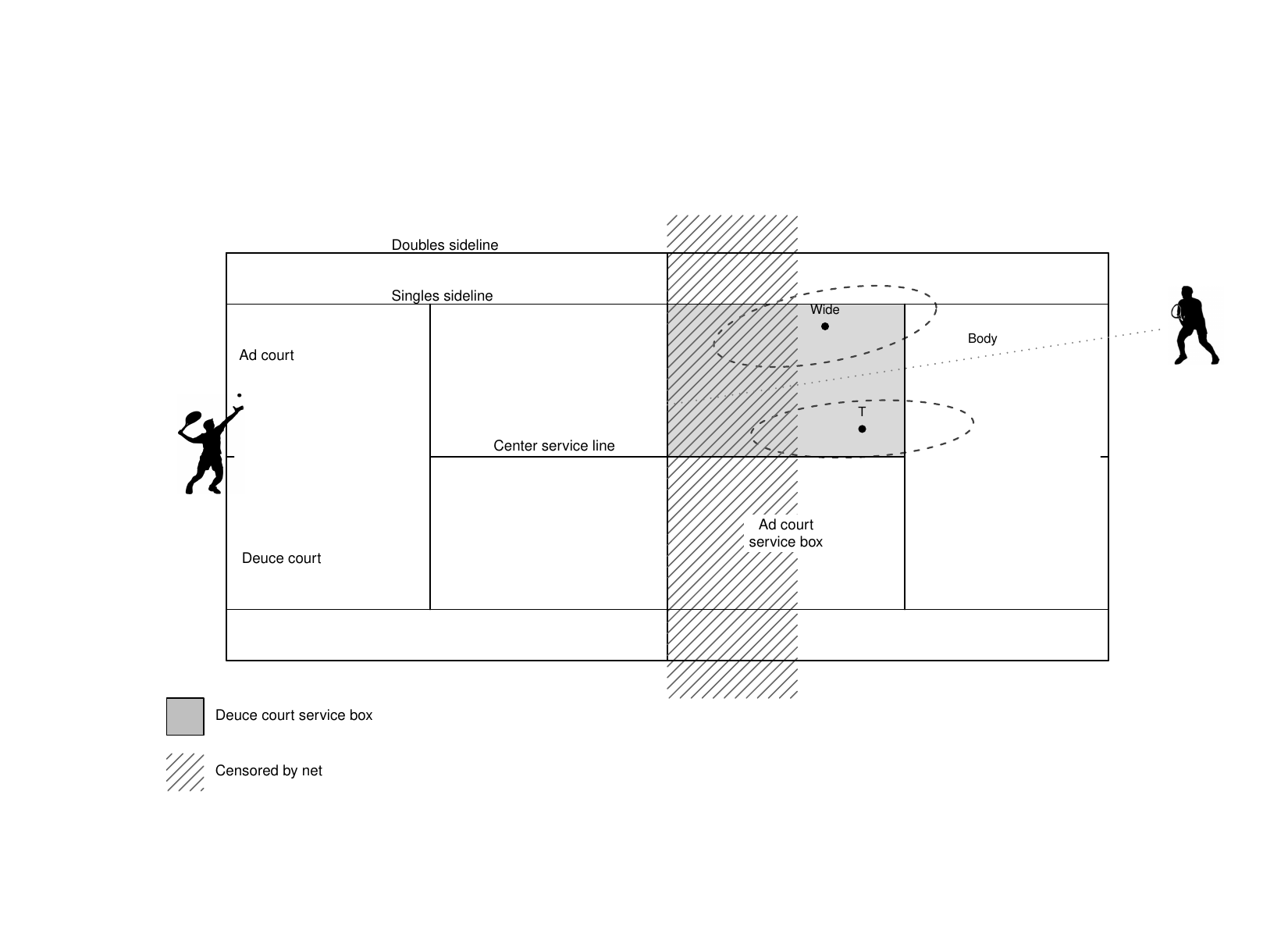}
    \caption{Illustration of the serving decision for a Deuce court serve.  The center service line divides each side of the net into the Deuce court, the half to a player's right as they face the net, and the Ad court, the half to their left.  Serves are struck diagonally, so a Deuce court serve is hit from the server's Deuce half into the returner's Deuce service box. Hypothetical targets for the server's Deuce court Wide and T serves are shown by dots, and the ellipse surrounding each target represents the server's execution error.  The dotted line traces a hypothetical body-serve trajectory toward the returner.  Diagonal strokes denote the approximate region for which serves will be censored (i.e., hit the net). 
    }
    \label{fig:intro_fig}
\end{figure}

Every point in tennis is initiated by a \textit{serve}, in which the serving player hits the ball to the returner on the opposite side of the court.  For a serve to be valid, or \textit{in}, it must land in the diagonally opposite \textit{service box} (see Figure \ref{fig:intro_fig}) without hitting the net.  
If the serve lands outside the service box, including instances where it hits the net and fails to cross over, it is called a \textit{fault}.  The server is then allowed one additional attempt. Failing to land the second serve in bounds results in a \textit{double fault}, and the returner wins the point. 

The serve is the most consequential shot in tennis, as it initiates every point and strongly influences point outcomes \citep{mecheri2016serve,montagna2021bayesian, o2008importance}.  It is also the only shot entirely under the player's control, since there is no incoming ball to react to.
The decision of where to aim the serve is therefore a key component of a player's strategy.  

As illustrated in Figure~\ref{fig:intro_fig} for a Deuce court serve, there are three serve-placement strategies in tennis: aiming ``out wide'' (Wide), which entails serving toward the singles-boundary side of the service box; aiming ``up the T'' (T), in which the server aims toward the center service line boundary of the service box; and aiming into the returner's body (Body), attempting to jam the returner between their forehand and backhand to force an awkward return.

In the cases of the Wide and T strategies, the server seeks to direct the ball away from the returner’s position on either their forehand or backhand side.  Because the returner must guard against both strategies, neither one dominates and the server benefits from mixing between them.  
This gives rise to a natural game-theoretic problem that has been studied extensively in the tennis analytics literature \citep{walker2001minimax, hsu2007minimax, gauriot2023expertise, anderson2025disequilibrium}. 

The within-region aiming decision, by contrast, has received comparatively little attention. Once a player has selected a serve strategy (Wide, Body, or T), the player must choose a specific aim location \textit{within} the selected region. For both Wide and T serves, the ideal locations for the ball to land lie along the horizontal boundaries of the service box, where the returner must travel the greatest distance to reach the serve. 
However, as illustrated by the dashed ellipses in Figure~\ref{fig:intro_fig}, serves are subject to \emph{execution error} which induces variability around the intended targets. This creates a trade-off for the server: aiming closer to the service-box boundaries shifts the serve distribution toward locations that are harder to return, but it also increases the risk of a fault. Conversely, hedging inward from the boundaries reduces the probability of faulting, but generally yields less favorable serve locations when the ball lands in play. While less pronounced, a similar tradeoff exists in how deep or shallow to aim the serve.  This paper addresses the resulting spatial decision of precisely where within a selected region a player should aim.  We restrict our focus to the Wide and T strategies; since a Body serve aims to jam the returner rather than to exploit proximity to a boundary, its aiming problem is comparatively insensitive to execution error.

How far a player should hedge in from a service-box boundary depends on the magnitude of their execution error.  Execution error, however, can be estimated only imperfectly from ordinary match-play data because it is defined relative to the player’s intended targets, which are unobserved. To address this limitation, we conducted an experiment with members of \if0\anon a Division~1 NCAA university's men's tennis team\else the Brigham Young University men's tennis team\fi{} in which the serve targets were known.  Prior to serving, each player explicitly identified the targets they believed to be optimal for their Wide and T serves from both the Ad and Deuce courts. These targets were marked on the court and were visible while they served.  Throughout the paper, we refer to these locations as the players' \emph{stated targets}, denoted by $\mathbf{m}$. Players then served repeatedly toward each target and the bounce locations were recorded.

Using these data, we estimate player-specific execution error distributions using a Bayesian hierarchical model that accounts for serves censored by net contact. Combining the estimated execution error distributions with a model for the value of landing a serve at a given location, we can determine where a player should optimally aim. We cast this as a two-period Markov decision process (MDP), where the decision variables are the first- and second-serve aim locations. We solve the MDP for each posterior draw of the execution error distributions, yielding a posterior distribution over each player’s optimal aim location, which we denote by $\mathbf{m}^*$.

In running this experiment, we discovered an unexpected phenomenon. Despite explicitly identifying their targets and having them visibly marked on the court, players’ serves often did not center around the locations they specified. 
This introduced a third quantity of interest beyond the stated and optimal targets: the \emph{realized centers} of the players' serve distributions, denoted by $\widetilde{\mathbf{m}}$. 
Eliciting players’ intended targets therefore did not eliminate the need to estimate the centers of their serve distributions. It did, however, allow us to compare players’ stated targets to their realized centers, which would not have been possible from observational match-play data alone.

\begin{figure}[H]
    \centering
    \includegraphics[trim = 0in .7in 0in .55in, clip, width=1\textwidth]{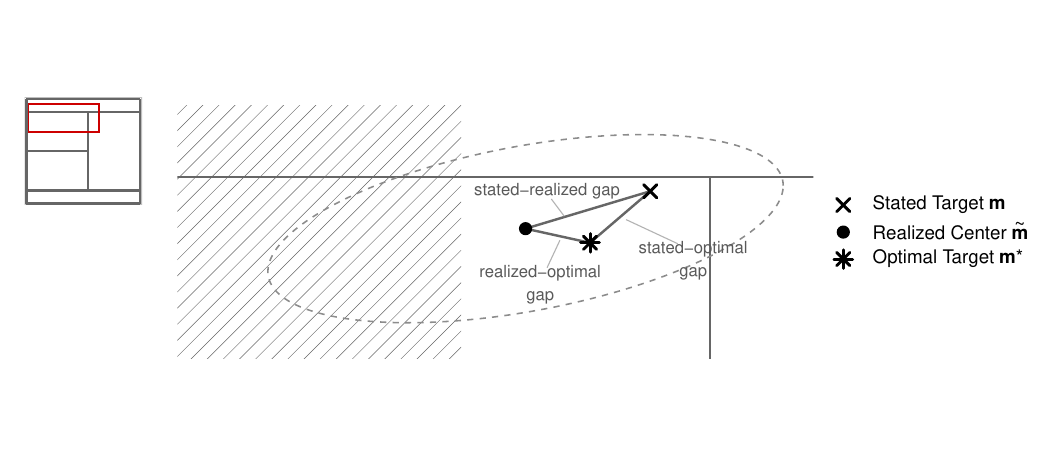}
    \caption{Conceptual illustration of $\mathbf{m}$, $\widetilde{\mathbf{m}}$, and $\mathbf{m}^*$, and their pairwise relationships, for the Deuce court Wide serve from Figure~\ref{fig:intro_fig}.  The serve distribution and net-censored region are as in Figure~\ref{fig:intro_fig}.
    }
    \label{fig:bias_concept}
\end{figure}

Comparing $\mathbf{m}$, $\widetilde{\mathbf{m}}$, and $\mathbf{m}^*$ sheds light on three related questions: 
\begin{enumerate}
    \item \emph{Stated--realized gap} ($\mathbf{m}$ vs.\ $\widetilde{\mathbf{m}}$): do players' stated targets match where they actually aim when serving?
    \item \emph{Stated--optimal gap} ($\mathbf{m}$ vs.\ $\mathbf{m}^*$): do players' stated targets align with the player-specific optima we estimate?
    \item \emph{Realized--optimal gap} ($\widetilde{\mathbf{m}}$ vs.\ $\mathbf{m}^*$): do players' realized centers align with the estimated optima?
\end{enumerate}
Figure~\ref{fig:bias_concept} illustrates these quantities and relationships.

We find strong evidence of a systematic gap between players’ stated targets and realized serve centers (i.e., stated-realized gap). In 57 of 64 player--serve period--region combinations, players' stated targets were more aggressive (i.e., closer to the service-box boundaries) than the realized centers.  This gap was largest for second serves. 

Interestingly, the realized centers were themselves systematically more aggressive than the estimated optima (in 44 of 64 combinations). However, they were substantially closer to the estimated optima than the stated targets were. Players thus articulate riskier targets than they play, and their observed behavior is more optimal than their stated targets.  

\paragraph*{Data and Code} 
All of the experiment data is publicly available at \url{https://github.com/nsandholtz/byu_tennis_serves}.  This repository also includes STAN code of the serve execution model and R code to reproduce the primary results in the paper. Output from \cite{kovalchik2020space}, which models a serve's expected point value conditional on its landing location, is also included, which is necessary to estimate optimal targets (see Section \ref{sec:value_surface}).  

\section{Related Work} \label{sec:related_work}

The optimal aiming of a tennis serve is an instance of the general motor-planning problem studied by \citet{trommershauser2003statistical}, who show through simple laboratory aiming tasks that the optimal place to aim depends on both the task payoff structure and the decision maker’s motor variability. Within sport, this framework has been developed most extensively in darts. \citet{tibshirani2011statistician} use data from darts aimed at the bullseye to fit a Gaussian model of execution error, then combine this model with the dartboard’s reward surface to identify the aim point that maximizes the expected score from a single throw. Subsequent work has extended this analysis to different data settings \citep{miller2021monte}, alternative execution-error models \citep{archibald2024estimating}, the computation of optimal strategy within the larger game \citep{haugh2022play,haugh2024empirical}, and handicapping \citep{chan2023no}.

Similar aiming problems have been studied in soccer and baseball. 
In soccer, \citet{hunter2018modeling, hunter2022identifying} conducted penalty-kick experiments in which players aimed at known targets and the resulting ball locations on the plane of the goal were recorded. They used these data to estimate bivariate Gaussian execution-error models of execution error, which they combined with a model of goalkeeper coverage to identify aim locations that maximize scoring probability.
\citet{baron2024miss} addressed the inverse problem, recovering information about a player's intended targets from their off-target shots in soccer matches.
In baseball, \citet{melville2023game} embedded a model of pitcher accuracy within a game-theoretic model of pitch sequencing to recommend pitch locations that account for each pitcher's execution distribution.  
\citet{ludwig2025xctrl} use  mixture models to estimate a pitcher's deviations from individualized target zones, which could be incorporated within a framework like that of \citet{melville2023game}.

In tennis, the serving decision has been studied predominantly as a game-theoretic problem over a discrete action space (e.g., Wide vs.\ T).
In this framework, \citet{walker2001minimax} found that professional players' serve-direction choices are broadly consistent with minimax play.
More recent work by \citet{anderson2025disequilibrium}, however, reaches the opposite conclusion. Testing the minimax condition across the entire service game rather than point by point, they posit that many elite professionals could significantly raise their win rates by adopting the best-response serve strategies they compute via dynamic programming.

A related tennis literature studies risk-taking across the two serves, modeling the ``aggressiveness'' of a serve by either the serve's speed or probability of landing in.
\citet{george1973optimal} and \citet{gerchak2017serving} mathematically show that the option value of the second serve should make the optimal first serve riskier than the second. 
Using data from Wimbledon, \citet{klaassen2009efficiency} estimate optimal fault probabilities for first and second serves. They find that professionals take more risk on first serves than on second serves, but could increase their probability of winning the point by serving more aggressively on second serves. \citet{ely2017agents} use ``let'' serves as a natural experiment, finding that players' risk allocation between first and second serves is generally consistent with optimality.\footnote{A serve that contacts the net but still lands in the service box is a \textit{let}, in which case no point is awarded and the serve is replayed.}

The work most closely related to ours is that of \citet{chan2022markov}. Using a coarse partition of the opponent's half court as their action space, they develop a Markov process model to recover optimal aiming strategy for an arbitrary shot in a tennis rally under execution error. Their approach, however, has an important limitation: in the absence of both experimental and observational shot location data, the execution-error distributions they construct are largely governed by the geometry of the action space partition rather than by observed serve behavior. The method is also limited in that it does not account for net censoring. 

We build directly on this work. Restricting our focus to the serve, we run an experiment that elicits players' intended targets and collect the precise bounce locations of the resulting serves.  These data allow us to accurately estimate execution-error distributions while accounting for serves censored by net contact.  By solving the MDP across posterior draws of the execution-error distributions, we also quantify uncertainty in each player’s optimal aim locations.

\section{Experiment Design and Data Collection} \label{sec:experiment}

We conducted an experiment with eight players on the \studyteam{}, an NCAA Division I athletic program, to investigate player execution error on serves.\footnote{This study was reviewed by the \studyirb{}, which determined that it does not meet the regulatory definition of human subjects research under 45~CFR~46.102(e)(1) and therefore was not subject to IRB oversight.}  
Players were first asked to identify what they believed to be their optimal aiming locations (i.e., their stated targets~$\mathbf{m}$) for four serve regions of interest: Ad Wide, Ad T, Deuce Wide, and Deuce T.  As the player identified each target location, a student assistant marked the spot on the court with silver duct tape in the shape of an ``X'', clearly visible from where they would be serving.
All players stated that their aiming locations would remain unchanged between serve periods. Some players were present during others’ target selection, which may have influenced their responses.

After warming up, the players iteratively hit serves to randomly assigned service courts (Ad or Deuce) under randomly assigned strategies (Wide or T).  Each serving sequence began with a first serve where the player served to a returner who was unaware of the assigned strategy.  They were instructed to aim at their chosen target for the assigned region and serve as if in a match-play setting. No restrictions were placed on where players stood along the baseline or on the type of serve they used (e.g., flat or kick).  The returner attempted to return each serve, but the server did not continue the rally after the return attempt. The experiment was therefore designed to measure serve execution rather than complete point outcomes.\footnote{We estimate a serve’s expected value conditional on its landing location by a separate model (see Section~\ref{sec:solving_optimal_aiming}).}  If the first serve was a fault, the player delivered a second serve, for which the court side remained the same but the strategy was randomly reselected. This process continued until each player had attempted approximately 70 serves. These sessions broadly resembled the players' usual serving practice.

To measure the bounce location of each serve, a research assistant lightly misted the ball with water before each serve, so that it left a visible but temporary mark where it landed. All players agreed that the lightly misted balls did not interfere with the serve quality.  Serve speeds were measured using a Stalker\textsuperscript{\tiny \textregistered} Solo 2 digital sports radar gun positioned directly behind the server.\footnote{This stationary Doppler radar gun has a speed range of 5--600 MPH and an accuracy of $\pm 3\%$ of the reading. The reported speeds reflect the scalar magnitude of the serve velocity vector at impact.}  For serves that crossed the net, a research assistant placed a numbered sticker at the bounce mark corresponding to the serve index. No spatial coordinates were recorded for serves that went into the net.

\begin{figure}[H]
    \centering
    \includegraphics[trim=0in .25in 0in .35in, clip, width=1\textwidth]{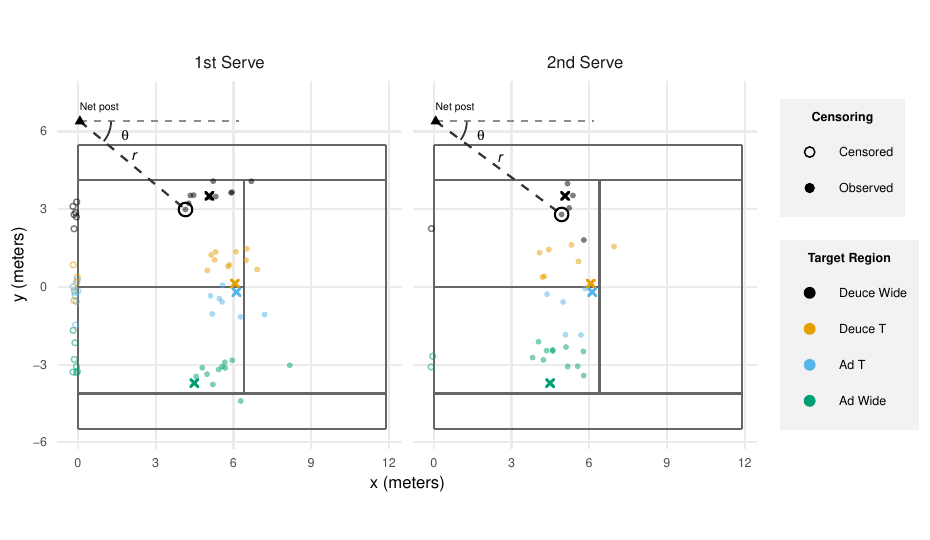}
    \caption{Observed data for Player 2 from the experiment.  X's denote the player's stated target locations, filled circles show observed bounce locations, and unfilled circles represent serves that hit the net (plotted along the net with vertical jitter for visibility).  
    In each panel, the polar measurement geometry is illustrated on a single representative serve.
    }
    \label{fig:tennis_data}
\end{figure}

After each session, bounce locations were recorded in polar coordinates using the net post on the left-hand side of the court as the reference anchor (see Figure \ref{fig:tennis_data}). A soft tape measure was used to record the distance $r$ of each sticker to the post.  The angle $\theta$ was measured using a laser digital angle finder, referenced to the line extending upward from the post and parallel to the left service line.  We subsequently converted these measurements to Cartesian coordinates, placing the origin at the intersection of the center service line and the net, via
\begin{align*}
x &= r\cos\theta + x_{\text{post}}, \\
y &= y_{\text{post}} - r\sin\theta,
\end{align*}
where $(x_{\text{post}}, y_{\text{post}}) = (0.076, 6.390)$ denotes the position in meters of the reference post in the Cartesian frame.
Figure~\ref{fig:tennis_data} displays the experiment data for Player 2, where the polar measurement geometry is illustrated on a representative serve in each panel. 
Observed data for all eight players are shown in Figure~\appref{fig:tennis_data_all}{C.1} in \appsec{sec:additional_figures}{C}. 

We collected 567 serves in total, for an average of 71 per player. Table~\appref{tab:sample_sizes}{C.1} in \appsec{sec:additional_figures}{C} reports the number of serve attempts for each player, serve period, target region combination. 
 Figure~\appref{fig:tennis_speeds}{C.2} in \appsec{sec:additional_figures}{C} shows histograms of serve speeds for all eight players, stratified by first vs second serve.  All players showed slower average speeds on second serves, which is consistent with match-play behavior. Finally, we note that all players were right-handed, except for Player 8.
\section{Modeling Serve Execution} \label{sec:execution_error}

In this section we develop a hierarchical Bayesian model of player execution error. We first specify a model of serve landing locations that accounts for censoring by the net, and then an interpolation scheme that extends the fitted model from the stated targets to arbitrary aiming locations. The resulting distributions supply the transition dynamics for the MDP developed in Section~\ref{sec:mdp}.

\subsection{Serve Execution Model}
\label{sec:serve_execution_model}

For a given player $p$, let $\mathbf{Z}^p_{ijk} = (Z^p_{x,ijk}, Z^p_{y,ijk})^\top$ denote the landing location of serve attempt $k = 1,\ldots,N^p_{ij}$ in serve period $i \in \{1,2\}$, aimed within region $j \in \{1,2,3,4\}$ (Deuce Wide, Deuce T, Ad T, and Ad Wide, respectively).\footnote{The number of serve attempts $N^p_{ij}$ per player, period, and region is reported in Table~\appref{tab:sample_sizes}{C.1} in \appsec{sec:additional_figures}{C}.}
We model $\mathbf{Z}^p_{ijk}$ as a bivariate Gaussian distribution,
\[
\mathbf{Z}^p_{ijk} \sim
\mathcal{N}_2\!\left(\widetilde{\mathbf{m}}^p_{ij},\, \boldsymbol{\Sigma}^p_{ij}\right),
\]
where $\widetilde{\mathbf{m}}^p_{ij}  \in \mathbb{R}^2$ is the unknown realized center of the serve distribution and $\boldsymbol{\Sigma}^p_{ij}$ is its covariance matrix. 
Importantly, \(\widetilde{\mathbf{m}}^p_{ij}\) is not constrained to equal the stated target \(\mathbf{m}^p_{ij}\).  

\subsubsection{Censoring}

We model the data using a Tobit-style censored normal model \citep{tobin1958estimation}, in which net contact acts as a censoring mechanism on the marginal distribution of $Z_x$. Specifically, we assume there exists an unknown threshold parameter $\alpha^p_{ij}$ in the horizontal ($x$) dimension such that serves with $Z^p_{x,ijk} \le \alpha^p_{ij}$ contact the net and do not produce an observable bounce location.  We accordingly define the net-clearance indicator $C^p_{ijk} := \mathbbm{1}\{Z^p_{x,ijk} > \alpha^p_{ij}\}$. For serves that clear the net ($C^p_{ijk} = 1$), the full two-dimensional landing location $\mathbf{Z}^p_{ijk}$ is observed; for serves that do not cross the net ($C^p_{ijk} = 0$), only the event $Z^p_{x,ijk} \le \alpha^p_{ij}$ is known. 

\subsubsection{Covariance structure}
We parameterize the covariance matrix \(\boldsymbol{\Sigma}^p_{ij}\) in terms of
marginal standard deviations and a correlation coefficient:
\begin{align}
   \boldsymbol{\Sigma}^p_{ij} =
\begin{pmatrix}
\tau^{p}_{x,ij} & 0 \\
0 & \tau^{p}_{y,ij}
\end{pmatrix}
\begin{pmatrix}
1 & \rho^{p}_{ij} \\
\rho^{p}_{ij} & 1
\end{pmatrix}
\begin{pmatrix}
\tau^{p}_{x,ij} & 0 \\
0 & \tau^{p}_{y,ij}
\end{pmatrix}, 
\label{eq:covariance_spec}
\end{align}
where \(\tau^{p}_{x,ij}\) and \(\tau^{p}_{y,ij}\) denote the marginal standard deviations in the horizontal and vertical directions, respectively, and \(\rho^{p}_{ij}\) governs the correlation between dimensions.  

\subsubsection{Likelihood}
Let \(\mathcal{C}^p_{ij}\) and \(\mathcal{H}^p_{ij}\)
denote the index sets of net-clearing and net-hitting serves,
respectively, for player \(p\) in serve period \(i\) aimed within region \(j\).
The likelihood contribution for the serves from a given \((p,i,j)\) combination is
\begin{align}
L^p_{ij}
&=
\prod_{k \in \mathcal{C}^p_{ij}}
\phi_2\!\left(
\mathbf{z}^p_{ijk}\,;\,
\widetilde{\mathbf{m}}^p_{ij},\,
\boldsymbol{\Sigma}^p_{ij}
\right)
\;\times\;
\prod_{k \in \mathcal{H}^p_{ij}}
\Pr\!\left(C^p_{ijk} = 0\right) 
\nonumber\\
&=
\prod_{k \in \mathcal{C}^p_{ij}}
\phi_2\!\left(
\mathbf{z}^p_{ijk}\,;\,
\widetilde{\mathbf{m}}^p_{ij},\,
\boldsymbol{\Sigma}^p_{ij}
\right)
\;\times\;
\left[
\Phi\!\left(
\alpha^p_{ij}\,;\,
\widetilde{m}^p_{x,ij},\,
\big(\tau^p_{x,ij}\big)^{2}
\right)
\right]^{H^p_{ij}},
\label{eq:full_censored_lik}
\end{align}
where \( H^p_{ij} = |\mathcal{H}^p_{ij}|\), \(\phi_2(\cdot\,;\boldsymbol{\mu},\boldsymbol{\Sigma})\) denotes the bivariate Gaussian density, and \(\Phi(\cdot\,;\mu,\sigma^2)\) denotes the univariate Gaussian CDF.  The full likelihood is the product of these contributions over all \((p,i,j)\):
\[
L \;=\; \prod_{p=1}^{8} \prod_{i=1}^{2} \prod_{j=1}^{4} L^p_{ij}.
\]

\subsection{Prior Specification} 
\label{sec:prior_specification}

We estimate all model parameters in a Bayesian framework.  With the exception of the location parameters $\widetilde{\mathbf{m}}$, we use hierarchical priors to share information across players, serve periods, or target regions.

\subsubsection{Location  ($\widetilde{\mathbf{m}}$)} \label{sec:prior_location}
Since the players were instructed to aim at their stated targets, we center each realized center prior at its corresponding stated target \( \mathbf{m}^p_{j} \), using an isotropic bivariate Gaussian distribution with global scale \( \sigma_\mu \):
\[
\widetilde{\mathbf{m}}^p_{ij} \sim \mathcal{N}_2(\mathbf{m}^p_{j}, \sigma_\mu^2 \mathbf{I}).
\]

\subsubsection{Scale ($\boldsymbol{\tau}$)} 
We expect that player-level characteristics are the main source of variation in execution-error magnitude, with greater differences across players than across cells within a player. We therefore shrink the cell-specific scales ($\boldsymbol{\tau}^p_{ij}$) toward player-level scales ($\boldsymbol{\tau}^p$):
\[
\log \tau^p_{x,ij} \sim \mathcal{N}(\log \tau^p_{x}, \sigma_\tau^2), \quad
\log \tau^p_{y,ij} \sim \mathcal{N}(\log \tau^p_{y}, \sigma_\tau^2).
\]
We assign lognormal priors to the player-level scales such that the prior medians correspond to 1.5m and 1.0m in the horizontal and vertical directions, respectively:
\[
\tau^{p}_{x} \sim \operatorname{LogNormal}\big(\log(1.5),\,0.75^2\big),\qquad
\tau^{p}_{y} \sim \operatorname{LogNormal}\big(\log(1.0),\,0.5^2\big).
\]
These values were chosen to reflect plausible magnitudes of serve execution error, with greater expected spread in the serve depths ($x$) than laterally ($y$).

\subsubsection{Correlation ($\rho$)} 
The orientation of the error ellipse is primarily governed by the angle between the impact location and the target. We therefore shrink $\rho^p_{ij}$ toward a region-level correlation. We model these on the Fisher-$z$ scale, which maps the constrained interval (-1,1) to the real line, permitting a Gaussian hierarchical specification:
\[
\zeta^p_{ij} \sim \mathcal{N}(\zeta_j, \sigma_\zeta^2), \qquad
\zeta_j \sim \mathcal{N}(0, 0.5^2),
\]
where $\zeta^p_{ij} := \operatorname{atanh}(\rho^p_{ij})$ and $\zeta_j := \operatorname{atanh}(\rho_j)$.

\subsubsection{Censoring ($\alpha$)} \label{sec:prior_censoring}
Net clearance depends on the arc of the serve, which is largely determined by its spin and speed. These features differ systematically by serve period because players typically hit flatter, faster first serves and use more topspin on second serves. We therefore shrink the cell-specific censoring parameters toward period-level means:
\[
\alpha^p_{ij} \sim \mathcal{N}(\alpha_i, \sigma_\alpha^2)
\;\; \text{truncated to } (0, a^p_{ij}),
\]
where the upper bound $a^p_{ij}$ is the minimum observed bounce depth ($x$-value) among player $p$’s serves in period $i$ and region $j$.  This reflects that the censoring threshold cannot exceed the shallowest serve actually observed. The lower bound of 0 ensures that the threshold lies beyond the net.  
We assign the period-level parameters $\alpha_i$ an informative prior,
$$
\alpha_i \sim \mathcal{N}(4, 1^2),
$$
reflecting prior knowledge that the minimum landing depth for serves clearing the net is around 4 meters beyond the net \citep{howard2004physics}.

Finally, we use weakly informative exponential priors for all global scale parameters:
\[
\sigma_\mu,\sigma_\alpha,\sigma_\tau, \sigma_{\zeta} \sim \text{Exponential}(5).
\]

\subsection{Model Validation and Fit} \label{sec:model_fit}

We tested the model described above against several alternative hierarchical structures using cross-validation. Our baseline was a nonhierarchical model in which each cell-level parameter was assigned an independent prior. Hierarchical pooling substantially improved predictive performance, which was expected given the relatively small number of serve attempts within each player--period--region combination (see Table~\appref{tab:sample_sizes}{C.1} in \appsec{sec:additional_figures}{C}). Pooling the execution-error scale parameters produced the largest gain in out-of-sample predictive performance. Full details of the cross-validation study are provided in Appendix~\appref{sec:model_validation}{A}\insupp.

\begin{figure}[H]
    \centering
    \includegraphics[trim = 0in .25in 0in .32in, clip, width=\textwidth]{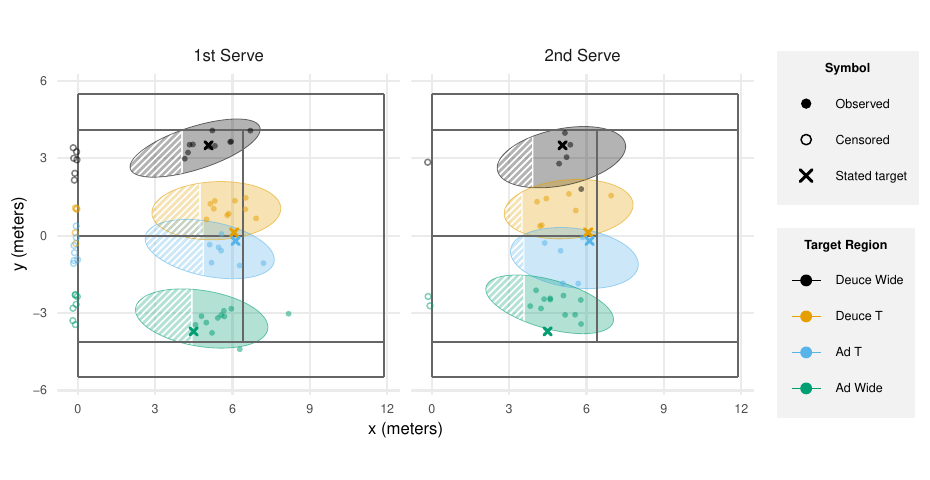}
    \caption{The fitted serve execution model for Player 2.  Shaded regions are 90\% contours of the fitted execution distribution $\mathcal{N}_2(\widetilde{\mathbf{m}}^p_{ij}, \boldsymbol{\Sigma}^p_{ij})$, evaluated at posterior means.  The hatched part of each contour shows the portion of the distribution below the posterior mean of the censoring threshold $\alpha^p_{ij}$.}
    \label{fig:exec_err_illustration}
\end{figure}

We fit the model via MCMC in Stan \citep{carpenter2017stan}. We ran four parallel chains with 24,500 iterations each, discarding the first 2,500 iterations of each chain as warmup. We set the chain length using the Raftery--Lewis diagnostic \citep{raftery1992practical}, which assesses the precision of posterior quantiles rather than the posterior mean. We applied this criterion to the 0.025 and 0.975 quantiles, requiring sufficient Monte Carlo precision that, with 95\% probability, the estimated mass in each corresponding tail was within 0.005 of the nominal value of 0.025. Convergence diagnostics indicated stable sampling, with the Gelman--Rubin diagnostic $\hat{R} < 1.001$ for all parameters and both bulk and tail effective sample sizes exceeding 12,000 for all parameters. Figure~\ref{fig:exec_err_illustration} illustrates the fit for Player 2.

We assessed the model fit through posterior predictive checks on the frequency of net (censored) serves and the distribution of serve landing locations.
To assess the model's fit of net serve frequency for each player-period-region combination, we computed 95\% posterior predictive intervals for the number of censored serves conditional on the total number of serve attempts in that cell. The observed counts were consistent with these intervals in all cases, as shown in Figure~\appref{fig:ppc_net}{C.3} in \appsec{sec:additional_figures}{C}.

To assess the model’s fit to observed serve locations, for each uncensored serve we generated a corresponding serve from its posterior predictive distribution using that serve's player, period, and region parameters.  Within each period–region combination, we then compared a range of quantiles of the simulated and observed landing locations in both the $x$ (depth) and $y$ (lateral) dimensions. We repeated this procedure 1,000 times to obtain posterior predictive intervals for each model-implied quantile. The resulting quantile-quantile plots are shown in Figure~\appref{fig:ppc_qq}{C.4} in \appsec{sec:additional_figures}{C}. 

In general, the observed quantiles fall within the model-implied 95\% intervals. The lateral dimension is well fit, with the exception of a minor departure from the Gaussian model for Ad Wide second serves.  
For second serves in the depth dimension, the observed quantiles tend to lie below the model-implied quantiles beyond the first quartile. This pattern suggests mild right skewness in serve depth that the Gaussian model cannot capture for these serves.

\subsection{Interpolation} \label{sec:interpolation}

Solving the optimal aiming problem (Section~\ref{sec:solving_optimal_aiming}) requires estimates of the serve execution distributions not only at the stated targets $\mathbf{m}$, but at all candidate aiming locations considered in the optimization.  We obtain these by linearly interpolating each player's covariance and censoring parameters between their T and Wide realized centers according to the angular position of the aiming location. We carry out this interpolation separately for each player, serve period, service court but we suppress these indices below for readability.

Let $\mathbf{a}$ denote an arbitrary
aiming location whose polar angle (measured from the serve impact location)
lies between those of the player’s T and Wide realized centers for either a Deuce or Ad court serve.
We linearly interpolate the covariance parameters ($\tau_x, \tau_y$, and $\rho$) and the censoring parameter ($\alpha$) via:
\[
\begin{aligned}
\tau_x(\mathbf{a}) &= (1 - w)\,\tau_{x,T} + w\,\tau_{x,W}, &\qquad
\tau_y(\mathbf{a}) &= (1 - w)\,\tau_{y,T} + w\,\tau_{y,W}, \\
\rho(\mathbf{a}) &= (1 - w)\,\rho_T + w\,\rho_W, &\qquad
\alpha(\mathbf{a}) &= (1 - w)\,\alpha_T + w\,\alpha_W,
\end{aligned}
\]
where \(w \in [0,1]\) represents the relative angular position of
\(\mathbf{a}\) between the player’s T and Wide realized centers (\(\widetilde{\mathbf{m}}_T\) and \(\widetilde{\mathbf{m}}_W\)), and the subscripts \(T\) and \(W\) denote the corresponding indices for the player’s T and Wide distributions, respectively. Specifically,
\[
w =
\frac{\theta(\mathbf{a}) - \theta(\widetilde{\mathbf{m}}_T)}
{\theta(\widetilde{\mathbf{m}}_W) - \theta(\widetilde{\mathbf{m}}_T)},
\]
where $\theta(\mathbf{a}) := \operatorname{atan2}(a_y - y_0, a_x - x_0)$ and where \((x_0,y_0)\) denotes the serve impact location, which we approximate by the center of the server's baseline.  If \(\mathbf{a}\) lies outside the angular range spanned by the
realized centers, we assign it the serve execution covariance parameters of the nearest
target (Wide or T).  Figure~\appref{fig:interp_illustration}{C.5} in \appsec{sec:additional_figures}{C} illustrates how the resulting execution-error ellipse morphs as the aiming location sweeps from the T anchor to the Wide anchor for a Deuce court serve.

\section{The Tennis Serve as a Markov Decision Process} \label{sec:mdp}

We now model a point in tennis as a two-period Markov decision process (MDP) in order to solve for optimal aiming locations.  For notational simplicity, we suppress player indexing throughout this section. 

An MDP is defined by a five-component tuple $\langle \mathcal{S}, \mathcal{A}, r, p, \gamma\rangle$, consisting of a state space $\mathcal{S}$, action space $\mathcal{A}$, reward function $r$, transition function $p$, and discount factor $\gamma \in [0, 1]$, governing how the values of future rewards diminish over time.  The objective of the decision maker is to identify a \textit{policy} $\pi: \mathcal{S} \to \mathcal{A}$ that maximizes the expected total discounted reward over the horizon $T$:
\[
\max_{\pi \in \Pi} \; \mathbb{E}_{\pi}\!\left[ \sum_{t=1}^{T} \gamma^{t-1} \, r(S_t, A_t) \right],
\]
where \(S_t \in \mathcal{S}\) and \(A_t \in \mathcal{A}\) are the state and action at time \(t\), respectively.

In order to treat the tennis serve as an MDP rather than a two-player game, we hold the returner's positioning fixed and absorb it into the reward function.  In our application, $T = 2$, corresponding to the first- and second-serve periods of a point.  The serve period is the only quantity that evolves within a point, so it is our sole state variable. We denote the state space as $\mathcal{S} = \{\text{first serve}, \text{second serve}\}$, with states indexed by $i \in \{1,2\}$.  Features of a point that influence the serving decision but remain fixed within an episode are conditioned on rather than modeled as state variables (e.g., identity of the returner or the court surface).  We collect these in a context vector $\mathbf{x}$, which in our experiment is simply the service court, Deuce or Ad, since all other contextual features were held fixed by design.  

The only possible state transition is from the first-serve state to the second, which occurs if and only if the first serve is a fault.\footnote{Because a let serve leaves the state of the point unchanged, we do not model it as a separate outcome.}  The transition dynamics are therefore entirely determined by the execution error model described in Section~\ref{sec:execution_error}.  Also, since winning a point on the second serve is just as valuable as winning on the first, we set $\gamma = 1$.  The remaining components of the MDP are described below.

\subsection{Actions} \label{sec:actions}

The decision variables in our formulation of the MDP are the spatial aiming locations of the first- and second-serve, which we denote by $\mathbf{a}_i \in \mathcal{A}$ for serve period $i \in \{1,2\}$.
The corresponding action space is $\mathcal{A}=\mathbb{R}^2$. 

While the action space is unrestricted in $\mathbb{R}^2$, the Wide and T strategies correspond to distinct regions of the court, and it will be useful to explicitly define them.  We draw two rays from the approximate serve impact location (the center of the server's baseline) through points located $0.5$ meters laterally to either side of the approximate returner position.\footnote{Returner locations were not recorded in our experiment. Instead, we approximated these locations using model output from \citet{kovalchik2020space}.} 
Beyond the net, these rays define three regions: the wedge between them is the Body region, while locations on the outward side of the Wide ray and the inward side of the T ray form the Wide and T regions, which we denote by $\mathcal{A}^{\text{W}}(\mathbf{x})$ and $\mathcal{A}^{\text{T}}(\mathbf{x})$, respectively.
We write $\mathcal{B}_{\text{Deuce}}$ and $\mathcal{B}_{\text{Ad}}$ for the Deuce and Ad service boxes, respectively, and more generally $\mathcal{B}(\mathbf{x})$ for the service box corresponding to context $\mathbf{x}$. 
Figure~\appref{fig:action_space}{C.6} in \appsec{sec:additional_figures}{C} illustrates the intersections of the strategy regions $\mathcal{A}^{\text{W}}(\mathbf{x})$ and $\mathcal{A}^{\text{T}}(\mathbf{x})$ with the service boxes $\mathcal{B}(\mathbf{x})$.

We note that while we only varied the aim location in the experiment, other variables could be treated as decision variables as well, including serve speed and spin.    Optimizing with respect to these variables is a promising direction of future work.

\subsection{Rewards}

The full reward structure in tennis is hierarchical, governing progression through points, games, sets, and matches. To simplify the analysis, we restrict attention to a single point episode. We define $R_i = 1$ if the server wins the point during serve period $i$, $R_i = -1$ if the server loses the point during serve period $i$, and $R_i = 0$ otherwise. For the first serve, $R_1 = 0$ corresponds to a fault; for the second serve, $R_2 = 0$ corresponds to a point that was already decided on the first serve (i.e., no fault on first serve). Since the point is decided in exactly one serve period, the total reward $R_1 + R_2 \in \{-1, 1\}$ encodes the point outcome.

 \subsection{The Optimal Aiming Problem} \label{sec:optimal_aiming_problem}

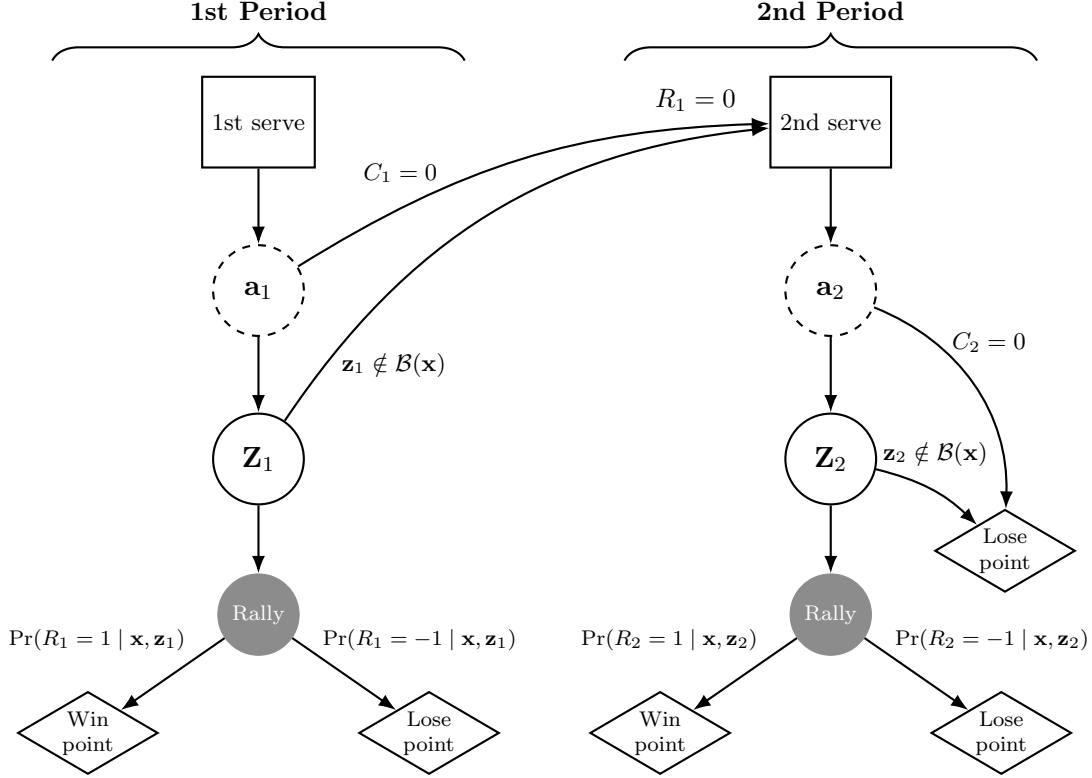
\begin{figure}[h]
    \centering
\begin{tikzpicture}[
  >=Latex,
  node distance=10mm and 26mm,
state/.style={draw, thick, rectangle, minimum size=12mm, align=center},
action/.style={draw, thick, dashed, circle, minimum size=12mm, align=center},
  obs/.style={draw, thick, circle, minimum size=12mm, align=center},
  terminal/.style={draw, thick, diamond, align=center, font=\scriptsize,
                   inner sep=1pt, aspect=1.7, minimum size=4mm},
  small/.style={font=\small},
  lbl/.style={font=\small, inner sep=1pt},
  outcome/.style={font=\small},
  br/.style={decorate, decoration={brace, amplitude=6pt}, thick}
]

\node[state] (s1) {\footnotesize{1st serve}};
\node[action, below=of s1] (m1) {$\mathbf{a}_1$};
\node[obs, below=of m1] (z1) {$\mathbf{Z}_1$};
\node[draw=gray!90, circle, fill=gray!90, text=white, font=\scriptsize,
      minimum size=10mm, align=center, below=9mm of z1]
      (rally1) {Rally};
\node[terminal, below left=9mm and 14mm of rally1] (w1) {Win\\point};
\node[terminal, below right=9mm and 14mm of rally1] (l1) {Lose\\point};
\draw[->, thick] (s1) -- (m1);
\draw[->, thick] (m1) -- (z1);

\draw[->, thick] (z1) -- (rally1);
\draw[->, thick] (rally1)  to[bend right=0] node[lbl, above, pos=.42, yshift=2mm,  xshift=-11mm,font=\scriptsize] {$\Pr(R_1 = 1\mid \mathbf{x},\mathbf{z}_1)$} (w1);
\draw[->, thick] (rally1)  to[bend left=0] node[lbl, above, pos=.42, yshift=2mm,  xshift=11mm,font=\scriptsize] {$\Pr(R_1 = -1\mid \mathbf{x},\mathbf{z}_1)$} (l1);

\node[state, right=60mm of s1] (s2) {\footnotesize{2nd serve}};
\node[lbl, left=4mm of s2, yshift=3mm] {$R_1=0$};
\node[action, below=of s2] (m2) {$\mathbf{a}_2$};
\node[obs, below=of m2] (z2) {$\mathbf{Z}_2$};
\node[terminal, below right=5mm and 14mm of z2] (fault2) {Lose\\point};

\node[draw=gray!90, circle, fill=gray!90, text=white, font=\scriptsize,
      minimum size=10mm, align=center, below=9mm of z2]
      (rally2) {Rally};

\node[terminal, below left=9mm and 14mm of rally2] (w2) {Win\\point};

\node[terminal, below right=9mm and 14mm of rally2] (l2) {Lose\\point};

\draw[->, thick] (m1) to[bend left=15] node[lbl, above, pos = .22, yshift=3mm, font=\footnotesize] {$C_1=0$} (s2);
\draw[->, thick]
  (z1) to[bend left=25]
  node[lbl, below, pos=.28, yshift=-8mm, font=\footnotesize]
  {$\mathbf{z}_1 \notin \mathcal{B}(\mathbf{x})$}
  (s2);
\draw[->, thick] (s2) -- (m2);
\draw[->, thick] (m2) -- (z2);

\draw[->, thick] (z2) -- (rally2);
\draw[->, thick] (rally2) to[bend right=0] node[lbl, above, pos=.42, yshift=2mm,  xshift=-11mm, font=\scriptsize]  {$\Pr(R_2 = 1\mid \mathbf{x},\mathbf{z}_2)$} (w2);
\draw[->, thick] (rally2) to[bend left=0] node[lbl, above, pos=.42, yshift=2mm,  xshift=11mm, font=\scriptsize]  {$\Pr(R_2 = -1\mid \mathbf{x},\mathbf{z}_2)$} (l2);

\draw[->, thick] (m2) to[bend left=35] node[lbl, above, yshift=4mm, xshift=2mm, font=\footnotesize] {$C_2=0$} (fault2);
\draw[->, thick]
  (z2) to[bend left=15]
  node[lbl, above, pos=.55, yshift=3mm, font=\footnotesize]
  {$\mathbf{z}_2 \notin \mathcal{B}(\mathbf{x})$}
  (fault2);
\node[small, above=6mm of s1] (p1lbl) {\textbf{1st Period}};
\node[small, above=6mm of s2] (p2lbl) {\textbf{2nd Period}};

\draw[decorate, decoration={brace, amplitude=10pt}, thick]
  ($(w1.north west |- p1lbl.south) + (0,-4mm)$)
  --
  ($(l1.north east |- p1lbl.south) + (0,-4mm)$);
\draw[decorate, decoration={brace, amplitude=10pt}, thick]
  ($(w2.north west |- p2lbl.south) + (0,-4mm)$)
  --
  ($(l2.north east |- p2lbl.south) + (0,-4mm)$);

\end{tikzpicture}     \caption{Illustration of a tennis point as a two-period MDP. Square nodes represent serve states (first and second serve). Unfilled circular nodes represent the server’s aiming locations and realized bounce locations. Filled circular nodes represent stochastic events (rallies), and diamond nodes represent terminal outcomes. Net faults (labeled $C_i=0$) result in censoring of the bounce location.
    }
    \label{fig:mdp_fig}
\end{figure}

Figure~\ref{fig:mdp_fig} provides a graphical summary of the MDP. The process begins in the 1st serve state, where the server selects aiming location $\mathbf{a}_{1} \in \mathcal{A}$.
On the second serve, the server selects a (possibly different) aiming location $\mathbf{a}_{2} \in \mathcal{A}$. In either period, the serve may result in an observed bounce location or a fault. A fault on the second serve results in an immediate loss of the point. In either period, conditional on the serve landing in play, the ensuing rally ends with the server winning or losing the point with probabilities $\Pr(R_i=1\mid\mathbf x,\mathbf z_i)$ and $\Pr(R_i=-1\mid\mathbf x,\mathbf z_i)$, respectively.

Given this structure, the basic formulation of the serving decision is the problem of choosing targets on first and second serve to maximize expected total reward over the episode:
\begin{align}
\label{eq:overall_problem}
\underset{\mathbf{a}_1, \mathbf{a}_2 \in \mathcal{A}}{\max}
\;\; \mathbb{E}[R_1 + R_2 \mid \mathbf{x}, \mathbf{a}_1, \mathbf{a}_2].
\end{align}
This formulation prescribes a single aiming location per serve period.  However, as discussed in Section \ref{sec:introduction}, effective serving requires mixing between the Wide and T strategies.  A server therefore needs a target for whichever strategy they choose on a given point, not only for the strategy that is better on average.
We therefore take as our object of interest not the maximizer of the expected-reward surface but its \emph{modes}. 

Let $Q_i$ denote the value of the point from the start of serve period $i$ as a function of the aiming location $\mathbf{a}_i$ in that period. Specifically,
\begin{align}
Q_2(\mathbf{a}_2 \mid \mathbf{x})
&:= \mathbb{E}[R_2 \mid \mathbf{x}, \mathbf{a}_2, R_1 = 0],
\label{eq:Q2_def} \\
Q_1(\mathbf{a}_1 \mid \mathbf{x}, \mathbf{a}_2)
&:= \mathbb{E}[R_1 + R_2 \mid \mathbf{x}, \mathbf{a}_1, \mathbf{a}_2] \nonumber \\
&\phantom{:}= \mathbb{E}[R_1 \mid \mathbf{x}, \mathbf{a}_1]
+ \Pr(R_1 = 0 \mid \mathbf{x}, \mathbf{a}_1)\, Q_2(\mathbf{a}_2 \mid \mathbf{x}),
\label{eq:Q1_def}
\end{align}
The second equality in \eqref{eq:Q1_def} relies on the fact that $R_2 = 0$ unless the first serve is a fault, and our assumption that execution error in one period does not depend on the aiming decision on the other period.  Consequently, $\mathbf{a}_2$ enters $Q_1$ only through $Q_2(\mathbf{a}_2 \mid \mathbf{x})$, allowing \eqref{eq:overall_problem} to be solved by backward induction.

Next, for an arbitrary function $h : \mathcal{A} \to \mathbb{R}$, let
\begin{align}
\label{eq:mode_set}
\mathcal{M}(h) := \bigl\{ \mathbf{a} \in \mathcal{A} \;:\; \exists\, \varepsilon > 0
\text{ such that } h(\mathbf{a}) > h(\mathbf{a}')
\text{ whenever } 0 < \lVert \mathbf{a}' - \mathbf{a} \rVert < \varepsilon \bigr\}
\end{align}
denote the set of strict local maxima, or modes, of $h$.  We can then define the optimal target for strategy $s \in \{\text{W}, \text{T}\}$ as the mode of the objective $Q_i$ within that strategy's region:
\begin{align}
\mathbf{m}_2^{*}(s \mid \mathbf{x})
&:= \mathcal{M}\bigl(Q_2(\cdot \mid \mathbf{x})\bigr) \cap \mathcal{A}^{s}(\mathbf{x}),
\label{eq:m2_star} \\
\mathbf{m}_1^{*}(s \mid \mathbf{x}, \mathbf{a}_2)
&:= \mathcal{M}\bigl(Q_1(\cdot \mid \mathbf{x}, \mathbf{a}_2)\bigr)
\cap \mathcal{A}^{s}(\mathbf{x}).
\label{eq:m1_star}
\end{align}
\section{Solving the Optimal Aiming Problem} \label{sec:solving_optimal_aiming}

Solving for the modes given by \eqref{eq:m2_star} and \eqref{eq:m1_star} requires estimates of two quantities: the serve execution distribution, and the value of a serve given its landing location. Writing the objectives (i.e., \eqref{eq:Q2_def} and \eqref{eq:Q1_def}) as integrals over the bounce location $\mathbf{z}$ makes this dependence clear:
\begin{align}
Q_2(\mathbf{a}_2 \mid \mathbf{x}) &= \iint_{\mathbb{R}^2}
f_2(\mathbf{z} \mid \mathbf{x}, \mathbf{a}_2)\,
V_2(\mathbf{z} \mid \mathbf{x}, \mathbf{a}_2) \, d\mathbf{z},
\label{eq:second_serve_ex_rew} \\
Q_1(\mathbf{a}_1 \mid \mathbf{x}, \mathbf{a}_2) &= \iint_{\mathbb{R}^2}
f_1(\mathbf{z} \mid \mathbf{x}, \mathbf{a}_1)\,
V_1(\mathbf{z} \mid \mathbf{x}, \mathbf{a}_1, \mathbf{a}_2) \, d\mathbf{z},
\label{eq:first_serve_ex_rew}
\end{align}
where $f_i$ is the execution error density of Section~\ref{sec:execution_error}, and $V_i$ is the value of a serve landing at $\mathbf{z}$ on the $i$th serve, which depends on whether the serve is in play or faults. Specifically,
\begin{align}
V_2(\mathbf{z} \mid \mathbf{x}, \mathbf{a}_2) &=
\begin{cases}
2\,\Pr(R_2=1 \mid \mathbf{x}, \mathbf{z}) - 1,
& \mathbf{z} \in \mathcal{B}(\mathbf{x}) \cap \ z_x > \alpha_2(\mathbf{a}_2), \\[3pt]
-1,
& \text{otherwise},
\end{cases}
\label{eq:V2_def} \\[6pt]
V_1(\mathbf{z} \mid \mathbf{x}, \mathbf{a}_1, \mathbf{a}_2) &=
\begin{cases}
2\,\Pr(R_1=1 \mid \mathbf{x}, \mathbf{z}) - 1,
& \mathbf{z} \in \mathcal{B}(\mathbf{x}) \cap \ z_x > \alpha_1(\mathbf{a}_1), \\[3pt]
Q_2(\mathbf{a}_2 \mid \mathbf{x}),
& \text{otherwise}.
\end{cases}
\label{eq:V1_def}
\end{align}
A serve is in play only if it clears the net and lands in the service box, which is reflected in the first case of both \eqref{eq:V2_def} and \eqref{eq:V1_def} above (recall that $\alpha_i(\mathbf{a}_i)$ is the interpolated censoring threshold defined in Section \ref{sec:interpolation}).  A fault on second serve is worth $-1$. A fault on first serve advances play to the second period, so it is worth the second serve objective itself, $Q_2(\mathbf{a}_2 \mid \mathbf{x})$.

While $f_i$ is estimated by the execution model of Section~\ref{sec:execution_error}, our experiment cannot inform $\Pr(R_i = 1 \mid \mathbf{x}, \mathbf{z})$, since the athletes did not play out the points they served.  We describe how we estimate these win probabilities next.

\subsection{Estimating Point-Win Probability} \label{sec:value_surface}

\begin{figure}[h]
    \centering
    \includegraphics[trim = .05in .3in .1in .4in, clip, width=1\textwidth]{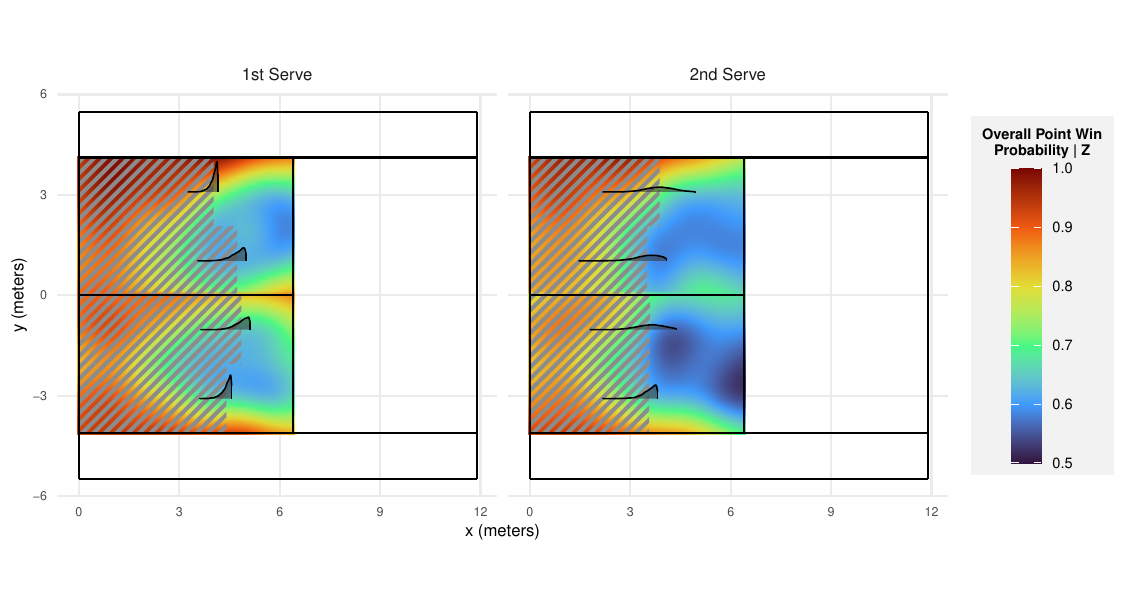}
    \caption{Estimated win probability surfaces for Player~2, conditioned on his average first- and second-serve speeds.  Gray-lined regions indicate the net-censored areas, defined using the posterior mean of $\alpha_{ij}^p$ for Player~2.  The shaded density within each region shows the full posterior of $\alpha_{ij}^p$.
    }
    \label{fig:value_function}
\end{figure}

Using ball-tracking data from multiple Australian Opens, \citet{kovalchik2020space} developed a generative model for the spatial trajectories of shots in a tennis rally, along with corresponding shot-level win probabilities. These models estimate the probability that an \textit{individual} shot ends the point, but they can be used recursively to estimate the overall probability that the server wins the point conditional on a serve landing at location $\mathbf z$. Following \citet{chan2022markov}, we approximate these point-level win probabilities via Monte Carlo simulation of rallies, conditioning on each player's average first- and second-serve speeds.  The procedure is explained in greater detail in Appendix~\appref{sec:reward_surface}{B}\insupp. The resulting surfaces for Player 2 are shown in Figure~\ref{fig:value_function}.  

We acknowledge that this outcome model was estimated from professionals rather than college athletes.  In the absence of collegiate tracking data, however, it provides a principled, data-driven approximation to the underlying reward surface.  Since the framework is modular, if collegiate ball-tracking data were available, the corresponding player-specific surfaces could be estimated directly and substituted without otherwise modifying our method.

Figure~\ref{fig:value_function} also depicts the portion of the service box that is censored by the net for each period-region combination.  Diagonal gray strokes indicate the net-censored areas, defined using the posterior mean of $\alpha_{ij}^p$ for Player~2.  The full posterior of each $\alpha_{ij}^p$ is shown by a miniature density plot centered within each region.  Note that the net-contact threshold is shallower on second serve than on first serve in each region for Player 2.  The same pattern holds across players.  Averaged over players and regions, $\alpha^p_{ij}$ sits $0.67$ meters closer to the net on second serve, and it is shallower in 31 of the 32 player-region combinations.  This is expected, since second serves are generally slower and carry heavier topspin than first serves.

\subsection{Estimating the Objective Surfaces} \label{sec:optimal_aiming_results}

To estimate the objective surfaces $Q_i(\mathbf{a}_i \mid \mathbf{x})$, we substitute posterior draws of the execution error parameters to the execution error densities $f_i(\mathbf{z}\mid\mathbf{x},\mathbf{a}_i)$ and the fixed point-win probabilities $\Pr(R_i=1 \mid \mathbf{x}, \mathbf{z})$ described above into \eqref{eq:second_serve_ex_rew} and \eqref{eq:first_serve_ex_rew}.  We then approximate the integrals numerically on a 0.1 meter grid over the relevant service court, evaluating $V_i$ as defined in \eqref{eq:V2_def} and \eqref{eq:V1_def}, with all remaining probability mass assigned the corresponding fault value.  This fault value is $-1$ on second serve. On first serve it is the continuation value $Q_2(\mathbf{a}_2 \mid \mathbf{x})$, which depends on the second-serve policy $\mathbf{a}_2$.   We assume the server selects the best available second-serve target, that is
\begin{align}
    \mathbf{a}_2^{*} := \argmax_{\mathbf{a}_2 \in \mathcal{M}(Q_2(\cdot \mid \mathbf{x}))} Q_2(\mathbf{a}_2 \mid \mathbf{x}),
    \label{eq:optimal_continuation}
\end{align}
and substitute $Q_2(\mathbf{a}_2^{*} \mid \mathbf{x})$ for the first-serve fault value in \eqref{eq:V1_def}.

\begin{figure}[h]
    \centering
    \begin{subfigure}{0.47\linewidth}
        \centering
        \includegraphics[trim=.25in 0 .39in 0 clip, width=\linewidth]{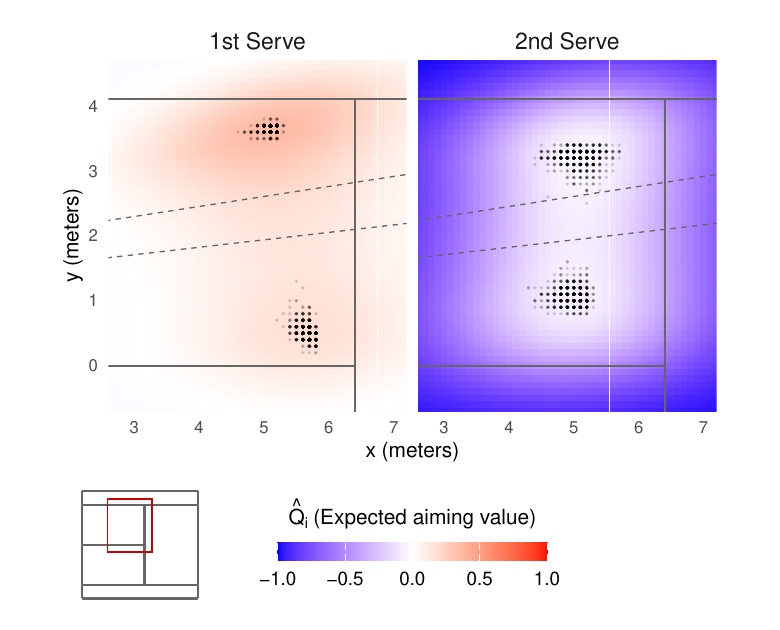}
        \caption{Deuce court.}
        \label{fig:optimal_aiming_a}
    \end{subfigure}
    \hfill
    \vrule width 0.5pt
    \hfill
    \begin{subfigure}{0.47\linewidth}
        \centering
        \includegraphics[trim=.25in 0 .39in 0 clip, width=\linewidth]{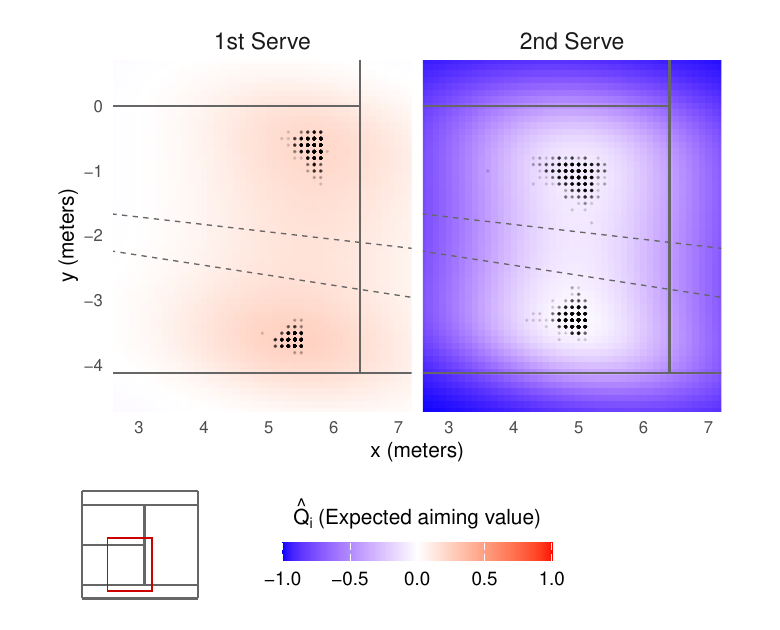}
        \caption{Ad court.}
        \label{fig:optimal_aiming_b}
    \end{subfigure}
    \caption{
    Estimated objective functions $Q_1(\mathbf{a}_1 \mid \mathbf{x}, \mathbf{a}_2^{*})$ and $Q_2(\mathbf{a}_2 \mid \mathbf{x})$ for Player 2. 
    Black dots denote posterior draws of the corresponding modes, $\mathcal{M}\bigl(Q_1(\cdot \mid \mathbf{x}, \mathbf{a}_2^{*})\bigr)$ and $\mathcal{M}\bigl(Q_2(\cdot \mid \mathbf{x})\bigr)$, for 1000 draws.  Dashed gray lines separate the Wide, Body, and T regions.
} \label{fig:optimal_aiming}
\end{figure}

Figure~\ref{fig:optimal_aiming} shows these estimated surfaces for Player 2, based on the posterior means of the serve execution model. The figure also displays 1000 posterior draws of $\mathcal{M}\bigl(Q_1(\cdot \mid \mathbf{x}, \mathbf{a}_2^{*})\bigr)$ and $\mathcal{M}\bigl(Q_2(\cdot \mid \mathbf{x})\bigr)$, the modes of the corresponding objectives for Player 2.  These dots represent our uncertainty about Player 2's optimal aiming locations for each strategy. Equivalent plots for all players are shown in Figures~\appref{fig:optimal_aiming_all_du}{C.7} and~\appref{fig:optimal_aiming_all_ad}{C.8} in \appsec{sec:additional_figures}{C}.

To approximate the posteriors of the first- and second-serve within-strategy optimal targets for a given strategy $s$ (i.e., $\mathbf{m}_1^{*}(s \mid \mathbf{x}, \mathbf{a}_2^{*})$ and $\mathbf{m}_2^{*}(s \mid \mathbf{x})$ from \eqref{eq:m2_star} and~\eqref{eq:m1_star}), we simply subset  the posterior draws of $\mathcal{M}\bigl(Q_1(\cdot \mid \mathbf{x}, \mathbf{a}_2^{*})\bigr)$ and $\mathcal{M}\bigl(Q_2(\cdot \mid \mathbf{x})\bigr)$ to those in $\mathcal{A}^{s}(\mathbf{x})$. We write $\widehat{\mathbf{m}}_{ij}^*$ for the point estimate of $\mathbf{m}_{ij}^*$, taken as the mean of the posterior draws of the mode that falls within region $j$ on serve period $i$.\footnote{With the second-serve strategy fixed at $\mathbf{a}_2^*$, the solutions to \eqref{eq:m2_star} and \eqref{eq:m1_star} form a family indexed by serve period $i \in \{1,2\}$ and region $j \in \{1,2,3,4\}$,  corresponding respectively to Deuce Wide, Deuce T, Ad T, and Ad Wide.}

\subsection{Properties of the Estimated Optima}

\subsubsection{Mode existence}

Two modes, one in the Wide region and one in the T region, do not arise for every posterior draw of the mode sets $\mathcal{M}\bigl(Q_1(\cdot \mid \mathbf{x}, \mathbf{a}_2^{*})\bigr)$ and $\mathcal{M}\bigl(Q_2(\cdot \mid \mathbf{x})\bigr)$. Where a region contains no mode, the objective climbs out of it toward the other strategy, so no aim within that region is locally optimal. Table~\appref{tab:mode_existence}{C.2} in \appsec{sec:additional_figures}{C} reports the proportion of posterior draws in which a mode existed for each player-period-region combination.

On first serves, a mode is present in nearly every posterior draw for the Deuce Wide, Ad Wide, and Ad T regions, with no player's proportion falling below 0.97. A mode in the first-serve Deuce~T region is similarly prevalent, but is less common for Player~1 (9\%) and Player~7 (83\%). On second serves, the absence of a mode becomes more common, particularly in the T regions. For example, six of the eight players were more likely to have no mode than to have one in the Deuce T region, and Player~6 had a mode in only 2\% of posterior draws. 

\subsubsection{Relative value of Wide and T optima}

The mode existence results suggest that the Wide and T modes may not be equally valuable in all cases. To examine this, Table~\appref{tab:mode_gap}{C.3} in \appsec{sec:additional_figures}{C} reports the value of the objective at each strategy's mode, $Q_i(\mathbf{m}_i^{*}(s \mid \mathbf{x}) \mid \mathbf{x})$ for $s \in \{W,T\}$, along with their difference (Wide $-$ T), averaged over draws in which both modes exist. A positive difference indicates that the Wide strategy has greater value.

On the Deuce court, the Wide mode is more valuable than the T mode for every player on both first and second serves. The average difference is $0.13$ on first serves and $0.03$ on second serves. On the Ad court, the average difference is $0.03$ on first serves, with the T mode more valuable for one player. On second serves, the average difference is approximately zero, with the T mode more valuable for four of the eight players.

These results help explain why we find fewer modes in the T region than in the Wide region. When the Wide strategy is sufficiently more valuable, the T region may no longer contain a distinct local mode. However, as discussed previously, both strategies remain strategically relevant even when one is weaker, since the value of serving Wide depends on the returner also needing to defend the T serve, and vice versa. The difference between the values of $Q_i$ at $\mathbf{m}_i^{*}(W \mid \mathbf{x})$ and $\mathbf{m}_i^{*}(T \mid \mathbf{x})$ therefore has implications for how a server should mix between the two strategies, which is beyond the scope of this paper.

\subsubsection{Optima can exist outside the service box} 

Because of serve execution error, the service-box boundary does not create a discontinuity in the aiming objective surface (see Figure \ref{fig:optimal_aiming}).  On second serve, $Q_2$ decays smoothly toward $-1$, the value of a double fault, as the aiming location moves outside the service box. On first serve, $Q_1$ instead decays toward the continuation value of the second serve, $Q_2(\mathbf{a}_2^{*} \mid \mathbf{x})$. 

This smoothness creates a rare but interesting phenomenon on first serve.  The optimal target can fall \emph{outside} the service box. Depending on the serve execution parameters and the win-probability surfaces, the continuation value can exceed the value of a poorly placed first serve by enough to pull the optimal target across the service-box boundary. In our posterior samples, this occurs exclusively in the Deuce T region on first serve.  
The optimal target crosses the center service line in 3.3\% of draws for Player~6, 2.5\% for Player~5, and 1.8\% for Player~4.
These effects are visible in the posterior draws of $\mathcal{M}\bigl(Q_1(\cdot \mid \mathbf{x}, \mathbf{a}_2^{*})\bigr)$ for these players in Figure~\appref{fig:optimal_aiming_all_du}{C.7} of \appsec{sec:additional_figures}{C}. 

\subsubsection{Second-serve optima are more conservative}

Without exception, the estimated second-serve modes lie further inside the service box from the relevant fault line (the singles sideline for the Wide strategy and the center service line for the T strategy) than the corresponding first-serve modes. Table~\ref{tab:hedge} reports the distance from each estimated mode to the relevant boundary, averaged across the eight players.
\begin{table}[H]
\centering
\begin{tabular}{lcccc}
\toprule
Serve period & Deuce Wide & Deuce T & Ad T & Ad Wide \\
\midrule
First  & 0.52 {\footnotesize\,(0.41, 0.72)} & 0.60 {\footnotesize\,(0.40, 1.09)} & 0.69 {\footnotesize\,(0.57, 0.93)} & 0.54 {\footnotesize\,(0.44, 0.73)} \\
Second & 0.98 {\footnotesize\,(0.82, 1.15)} & 1.26 {\footnotesize\,(0.99, 1.44)} & 1.14 {\footnotesize\,(0.98, 1.36)} & 0.99 {\footnotesize\,(0.83, 1.18)} \\
\bottomrule
\end{tabular}
 \caption{Average distance in meters from the strategy's estimated mode $\widehat{\mathbf{m}}_i^{*}(s \mid \mathbf{x})$ to the relevant lateral fault line: the singles sideline for the Wide strategy and the center service line for the T strategy. Parentheses give the range across the eight players. Larger values indicate a more conservative target.  
}
\label{tab:hedge}
\end{table}

Across all four regions, the second-serve target lies about twice as far inside the relevant boundary as the first-serve target. This pattern is consistent with \citet{gerchak2017serving}, who show theoretically that the second serve should have a strictly lower fault probability than the first serve and therefore be more conservative.
\section{Comparing Stated, Realized, and Optimal Targets} \label{sec:bias_analyses}

\begin{figure}[H]
    \centering
    \includegraphics[trim = .1in .1in .1in .1in, clip, width=1\textwidth]{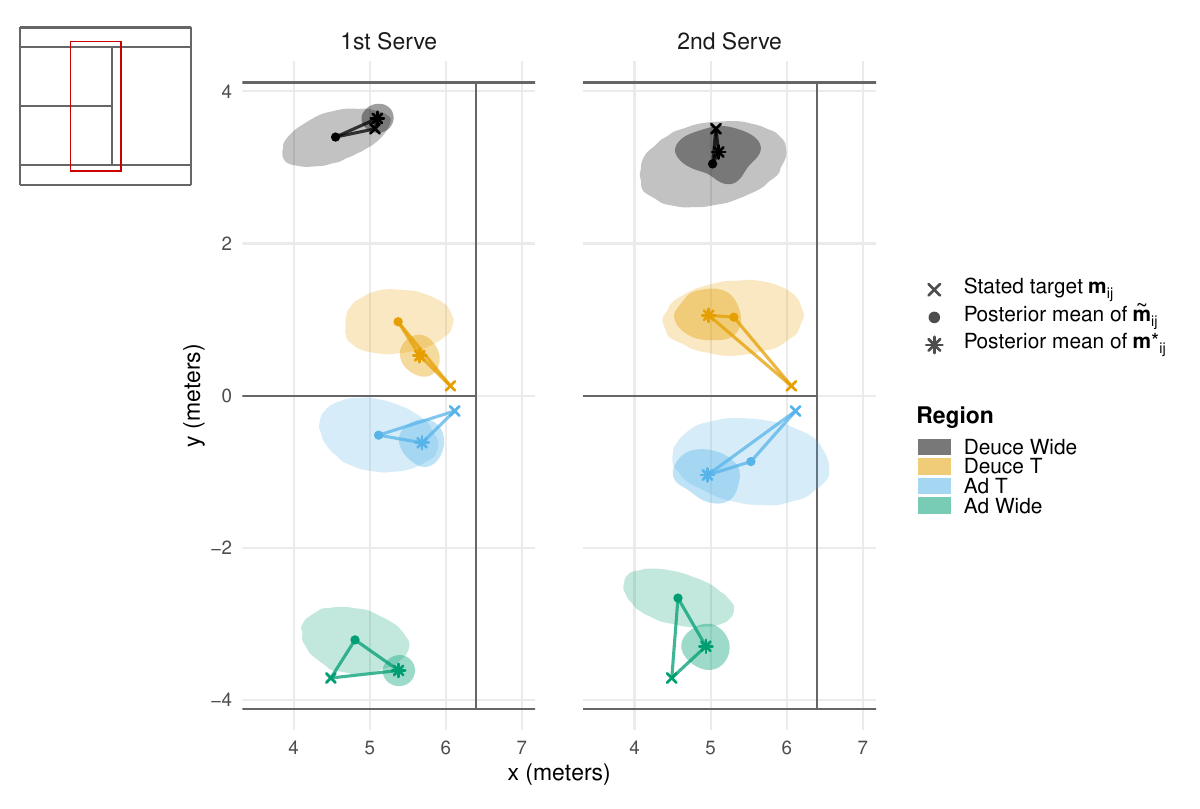}
    \caption{Depiction of the three aim comparisons for Player 2 on first serves (left panel) and second serves (right panel).    Lightly shaded regions show the 95\% highest-density region (HDR) of the posterior of $\widetilde{\mathbf{m}}_{ij}$; darker shaded regions show the 95\% HDR of the posterior of $\mathbf{m}^*_{ij}$. 
    }
    \label{fig:bias_triangle}
\end{figure}

With posterior distributions for both the realized centers (Section~\ref{sec:execution_error}) and the optimal aiming locations (Section~\ref{sec:solving_optimal_aiming}), we can compare all three pairings of the stated targets ($\mathbf{m}_{ij}$), realized centers ($\widetilde{\mathbf{m}}_{ij}$), and optimal targets ($\mathbf{m}^*_{ij}$).
Figure~\ref{fig:bias_triangle} depicts these comparisons for Player 2 in the context of the uncertainty on both $\widetilde{\mathbf{m}}_{ij}$ and $\mathbf{m}^*_{ij}$. The following subsections examine each comparison in turn.

\subsection{Stated--Realized Gap}

The \emph{stated--realized gap} is the difference between where a player said they were aiming ($\mathbf{m}_{ij}$) and where their serves actually centered ($\widetilde{\mathbf{m}}_{ij}$).  In Figure~\ref{fig:bias_triangle}, these gaps are the edges joining the Xs to filled circles.
In all eight of Player 2's region--period combinations, the estimated realized center of his serves lies \emph{inward} of his stated target, that is, farther from the strategy's lateral fault line (the singles sideline for Wide serves and the center service line for T serves).
In six of the combinations, the stated target falls outside the 95\% highest-density region of $\widetilde{\mathbf{m}}_{ij}$, indicating that the displacement is credibly nonzero.

\begin{figure}[H]
    \centering
    \includegraphics[width=0.95\textwidth]{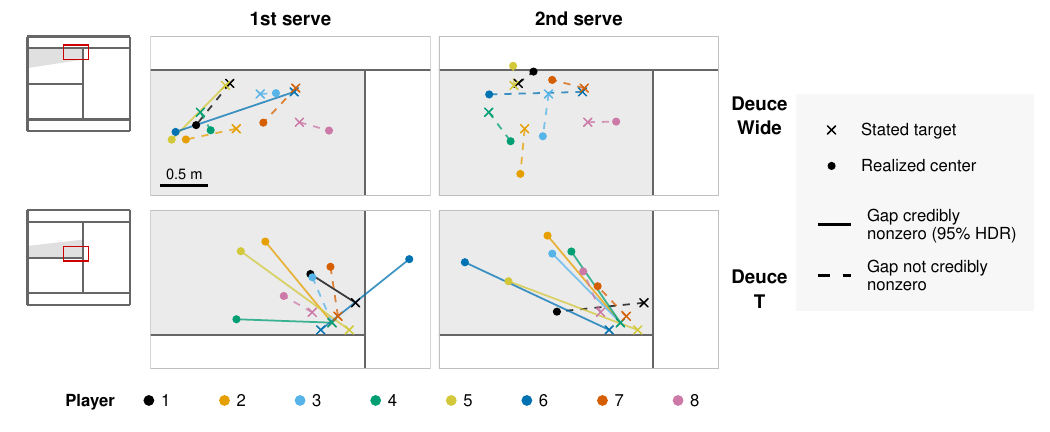}
    \caption{Stated--realized gap for all eight players on the Deuce court, shown in court space.  For each player, a segment joins the stated target $\mathbf{m}_{ij}$ (X) to the posterior mean of the realized serve-distribution center $\widetilde{\mathbf{m}}_{ij}$ (filled circle).  Solid segments mark gaps that are credibly nonzero at the 95\% level, meaning the stated target lies outside the 95\% highest-density region of the posterior of $\widetilde{\mathbf{m}}_{ij}$; dashed segments mark those that are not.  The test is on the two-dimensional displacement.
    }
    \label{fig:bias_court_deuce_stated_realized}
\end{figure}

This pattern is not unique to Player 2.  Figure~\ref{fig:bias_court_deuce_stated_realized} shows the stated--realized gap for all player--period--region combinations in the Deuce court. The Ad-court comparisons are shown in Figure~\appref{fig:bias_court_stated_realized}{C.9} in \appsec{sec:additional_figures}{C}.
In 56 of the 64 player--period--region combinations (88\%) the realized center hedges inward (i.e., away from the relevant strategy's lateral boundary) from the stated target.\footnote{The counts we reference throughout this section are totaled over both service courts and are hence out of 64.}   
 Of these 56, the stated target falls outside the 95\% highest-density region of the posterior of $\widetilde{\mathbf{m}}_{ij}$ in 32 (57\%).
 The inward hedge is larger on second serve in the Deuce T, Ad T, and Ad Wide regions (in the Deuce Wide region it is the same in both periods).  Averaging over the inward-hedged combinations, the mean lateral displacement grows from 0.37 meters on first serve (29 combinations) to 0.56 meters on second (27 combinations).  This is notable given that players uniformly stated their aiming locations would not change between serve periods. 

Players' realized aims were also systematically shallower than their stated targets.  In 51 of 64 combinations (80\%), the posterior mean of the realized center lies closer to the net than the stated target (i.e., $\widetilde{m}_x < m_x$).  The shallowness is concentrated in the T regions, where 31 of the 32 T combinations are shallower, by an average of 0.59 meters, versus 20 of the 32 Wide combinations by an average of 0.46 meters.  Unlike the lateral hedge, the depth bias does not grow between serve periods, averaging 0.54 meters on first serve and 0.54 meters on second, over the 26 and 25 shallower combinations, respectively.

We note that in both the case of the shallowness gap and the conservative lateral gap, our estimates are conservative due to our choice of priors on the location means of serve execution model.  Recall that we centered the prior for each realized center $\widetilde{\mathbf{m}}_{ij}$ at the stated target $\mathbf{m}_{ij}$, which shrinks the estimated realized center toward the stated target and thus biases \textit{against} finding a stated--realized gap.  The systematic gaps we recover therefore emerge despite a prior that pulls toward zero gap.

One possible explanation for this phenomenon is \textit{impression management}, which refers to the conscious or subconscious process in which people attempt to influence the other's perceptions by regulating and controlling information in social interaction \citep{sanaria2016conceptual, goffman1978presentation}.\footnote{for empirical examples of this effect, see \citet{ferris1984goal} and \citet{brandts2021gender}.}
In our study, players selected their targets in the presence of an assistant coach and, in some cases, one or two of their teammates.  It is plausible that they set more ambitious targets (perhaps subconsciously) to influence their coach's and peers' perceptions of them.

\subsection{Stated--Optimal Gap}

\begin{figure}[H]
    \centering
    \includegraphics[width=0.95\textwidth]{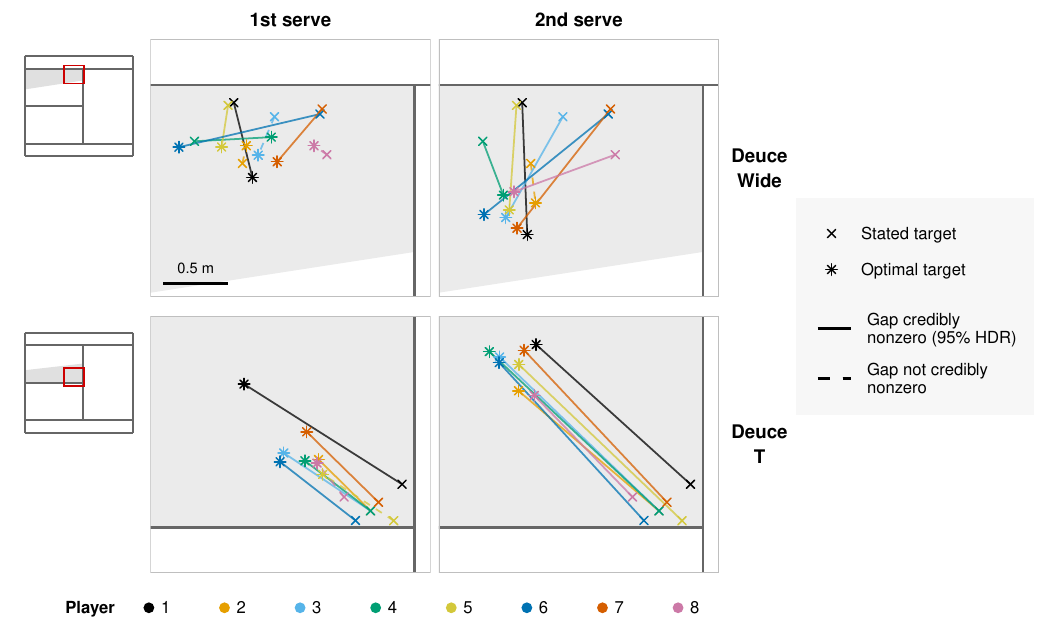}
    \caption{Stated--optimal gap for all eight players on the Deuce court.  Segments join each stated target $\mathbf{m}_{ij}$ (X) to the estimated optimal aiming location $\mathbf{m}^*_{ij}$ (asterisk), and are solid where the stated target lies outside the 95\% highest-density region of the posterior of $\mathbf{m}^*_{ij}$ over draws with an interior optimum, dashed otherwise.}
    \label{fig:bias_court_deuce_stated_optimal}
\end{figure}

The \emph{stated--optimal gap} is the displacement between a player's stated target $\mathbf{m}_{ij}$ and their optimal target $\mathbf{m}^*_{ij}$, shown by the edges joining Xs to asterisks in Figure~\ref{fig:bias_triangle}.
Figure~\ref{fig:bias_court_deuce_stated_optimal} compares all players' stated targets to their estimated optimal targets for Deuce court.   Ad-court comparisons are shown in Figure~\appref{fig:bias_court_stated_optimal}{C.10} in \appsec{sec:additional_figures}{C}.

The estimated optima lie inward (more conservative) of the stated targets in 61 of the 64 player--period--region combinations, 56 of which fall outside the 95\% highest-density region of the posterior of $\mathbf{m}^*_{ij}$.  The gap widens sharply between serve periods.  Averaging over the combinations whose optimum lies inward, the lateral displacement grows from 0.40 meters on first serve (29 combinations) to 0.85 meters on second (32).  On first serve this nearly matches the realized inward hedge of 0.37 meters, but on second serve it is well beyond the realized hedge of 0.56 meters.

The optima are also shallower than the stated targets in 52 of the 64 combinations (81\%).  As with the stated--realized gap, the shallowness concentrates in the T regions, where all 32 T combinations are shallower, by an average of 0.86 meters. For Wide combinations, 20 of the 32 are shallower, by an average of 0.48 meters.  In this case, the depth gap does grow between serve periods, from 0.52 meters on first serve to 0.91 meters on second.

Both the stated--realized and stated--optimal gaps therefore point in the same direction: players’ stated targets are more aggressive than their realized serve centers and, to an even greater extent, their estimated optimal locations. Both gaps are largest on second serves, where the realized centers lie an average of 0.56 meters inward of the stated targets and the estimated optima lie 0.85 meters inward.

\subsection{Realized--Optimal Gap}

\begin{figure}[H]
    \centering
    \includegraphics[width=0.95\textwidth]{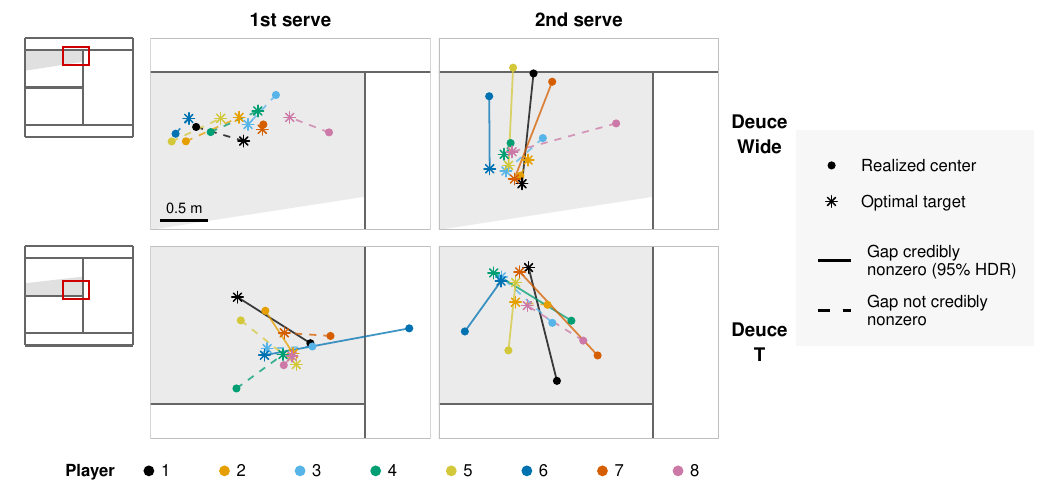}
    \caption{Realized--optimal gap for all eight players on the Deuce court.  Segments join each realized center $\widetilde{\mathbf{m}}_{ij}$ (filled circle) to the estimated optimal aiming location $\mathbf{m}^*_{ij}$ (asterisk).  A segment is solid where the origin falls outside the 95\% highest-density region of the posterior of the draw-paired displacement $\mathbf{m}^*_{ij} - \widetilde{\mathbf{m}}_{ij}$, dashed otherwise.}
    \label{fig:bias_court_deuce_realized_optimal}
\end{figure}

The \emph{realized--optimal gap} is the displacement between the center of a player's realized serve distribution ($\widetilde{\mathbf{m}}_{ij}$) and their estimated optimal aiming location ($\mathbf{m}^*_{ij}$), shown by the edges joining filled circles to asterisks in Figure~\ref{fig:bias_triangle}.  
Figure~\ref{fig:bias_court_deuce_realized_optimal} shows these comparisons for all players' Deuce court regions. The corresponding Ad-court comparisons are shown in Figure~\appref{fig:bias_court_realized_optimal}{C.11} in \appsec{sec:additional_figures}{C}.

The realized--optimal displacements are smaller than the other gaps on average.  The optimum lies inward of the realized center in 44 of the 64 player--period--region combinations, but the displacement is credibly nonzero in only 17 of 64.  Because both endpoints of this comparison are estimated, credibility is assessed on the displacement itself: pairing $\mathbf{m}^*_{ij}$ and $\widetilde{\mathbf{m}}_{ij}$ by posterior draw, we form the posterior of $\mathbf{m}^*_{ij} - \widetilde{\mathbf{m}}_{ij}$ and test whether the origin falls outside its 95\% highest-density region.  

Together, these results indicate that although players' \emph{stated} targets were credibly more aggressive than optimal in most cells, their \emph{realized} aiming behavior is not credibly different from optimal in the large majority of cells.  This may partly reflect differences between explicit and implicit decision-making \citep{keysers2008explicit,raab2003decision, engel2008better}. Experienced players may rely on intuitive, implicitly learned processes when selecting targets during play, and these processes may be imperfectly captured when players are asked to explicitly report an intended aim location. In the serving task, those familiar processes may have shifted their realized serve distributions toward more conservative and more optimal locations.

\section{Conclusion}
\label{sec:conclusion}

This paper set out to answer a deceptively simple question---where should a tennis player aim their serve?---and, in doing so, to assess how well players' beliefs and behavior align with what is optimal for them.  Using experimental data with known intended targets, we estimated each player's execution error, embedded the serving decision in a two-period MDP, and solved for player-specific optimal aiming locations while propagating posterior uncertainty throughout.  

We found that players’ stated targets were systematically more aggressive than optimal, but that their realized serve centers tended to lie inward of those targets, substantially narrowing the gap with the estimated optima. Thus, players articulated aggressive targets that they did not fully pursue, and their actual aiming behavior was much closer to optimal.

A natural extension is to incorporate spatial targeting into a fuller game-theoretic model of server--returner interaction, including optimal mixing across serve strategies. Such a model would require spatial data on returner positioning.  Finally, in ongoing work, we are applying these methods to observational data from professional tennis players to test whether realized--optimal gaps appear in professional match-play settings.

\section*{Acknowledgments}

We thank the members of the BYU men's tennis team for their participation and collaboration. We are grateful to Evan Miller, Tyler Duke, Xela Marchant, and Spencer Young for their assistance with data collection, and to Brett Pedersen, Chris Archibald, and Matt Heaton for helpful discussions and comments on the analysis and manuscript.

\bibliographystyle{plainnat}
\bibliography{bibliography.bib}

\newpage
\appendix
\counterwithin{figure}{section}
\counterwithin{table}{section}
\section{Serve Execution Model Validation} \label{sec:model_validation}

To evaluate and select among alternative serve-execution models, we compared eight candidate specifications using cross-validation. The candidates only differed in their hierarchical priors for the scales $\boldsymbol{\tau}^p_{ij}$, (Fisher-transformed) correlations $\zeta^p_{ij}$, and censoring thresholds $\alpha^p_{ij}$.  We did not vary the prior for the realized centers $\widetilde{\mathbf{m}}^p_{ij}$ because each player’s serve center is naturally expected to lie near the corresponding stated target, as reflected by the prior in Section~\mainref{sec:prior_location}{4.2.1}. We describe the eight candidate models below. Unless otherwise specified, each parameter retains the prior defined in Section~\mainref{sec:prior_specification}{4.2}.

\subsection*{Baseline Model}

The baseline model does not pool the scales, correlations, or censoring thresholds across cells. Instead, it assigns an independent prior to each cell-specific parameter:
\begin{align}
\log \tau^{p}_{x,ij} &\sim \mathcal{N}\!\left(\log 1.5,\, 0.75^2\right), \label{eq:baseline_scale} \\
\log \tau^{p}_{y,ij} &\sim \mathcal{N}\!\left(\log 1.0,\, 0.5^2\right), \nonumber  \\
\zeta^{p}_{ij} &\sim \mathcal{N}\!\left(0,\, 0.5^2\right), \label{eq:baseline_corr} \\
\alpha^{p}_{ij} &\sim \mathcal{N}\!\left(4,\, 1^2\right) \text{ on } (0, a^p_{ij}), \label{eq:baseline_cens}
\end{align}
where $\zeta^{p}_{ij} := \operatorname{atanh}(\rho^{p}_{ij})$ and $a^p_{ij}$ is the upper bound defined in Section~\mainref{sec:prior_censoring}{4.2.4}. 

\subsection*{Candidate Models}

The remaining seven models replace one or more of the independent priors defined above with the hierarchical priors in Table~\ref{tab:candidate_models}.  

\begin{table}[H]
    \centering
    \begin{tabular}{@{}llll@{}}
    \toprule
    Parameter & Anchor & Hierarchical prior  \\
    \midrule
    $\boldsymbol{\tau}^p_{ij}$
      & $\boldsymbol{\tau}^p$ (player)
      & $\log \tau^{p}_{x,ij} \sim \mathcal{N}(\log \tau^{p}_{x},\, \sigma_\tau^2)$ \\
      & 
      & $\log \tau^{p}_{y,ij} \sim \mathcal{N}(\log \tau^{p}_{y},\, \sigma_\tau^2)$ \\
    $\zeta^{p}_{ij}$
      & $\zeta_j$ (region)
      & $\zeta^{p}_{ij} \sim \mathcal{N}(\zeta_j,\, \sigma_\zeta^2)$ \\
    $\alpha^{p}_{ij}$
      & $\alpha_i$ (serve period)
      & $\alpha^{p}_{ij} \sim \mathcal{N}(\alpha_i,\, \sigma_\alpha^2)$ \\
    \bottomrule
    \end{tabular}
    \caption{Candidate hierarchical priors for $\boldsymbol{\tau}^p_{ij}$, $\zeta^p_{ij}$, and $\alpha^p_{ij}$.}
    \label{tab:candidate_models}
\end{table}
When a hierarchical prior is applied, the anchor takes on the corresponding independent prior defined in \eqref{eq:baseline_scale}-\eqref{eq:baseline_cens} (e.g., when the hierarchical prior is applied on the correlations, $\zeta_j \sim \mathcal{N}(0,\, 0.5^2)$).  We retain the truncation of $\alpha^{p}_{ij}$ to $(0, a^p_{ij})$ throughout. The remaining seven models represent all possible combinations in which hierarchical pooling is applied to at least one of the three parameter groups.

\subsection*{Cross-Validation Results}

\begin{figure}[H]
    \centering
    \includegraphics[trim = .1in .02in .4in .2in, clip, width=0.95\textwidth]{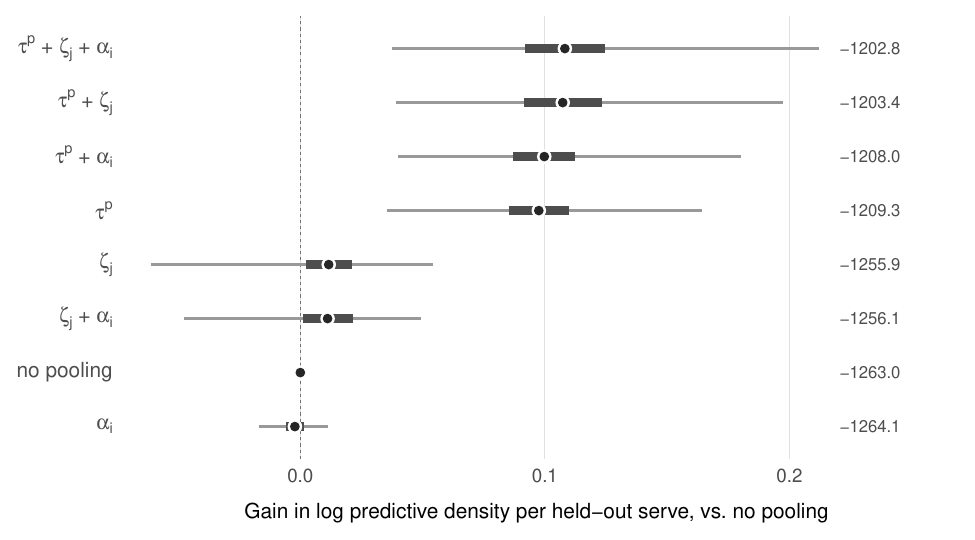}
    \caption{Gain in log predictive density per held-out serve relative to the baseline for the eight candidate models.  Rows are labeled by which parameters receive hierarchical priors. Each point marks the mean gain across the ten folds, the thick segment spans $\pm 1$ standard error of that mean, and the thin gray segment denotes the range of fold-specific gains. The total log predictive density, summed over all held-out serves, is shown at the right of each row.}
    \label{fig:factorial_elpd}
\end{figure}

We evaluated the eight models by ten-fold cross-validation.  Folds were stratified by player, serve period, region, and censoring status, so that each fold held out a small number of serves from a broad cross-section of cells.  We used the same folds for all models and computed standard errors from the per-serve differences in predictive performance.  The upper bound $a^p_{ij}$ was recomputed from the training rows of each fold.  Figure~\ref{fig:factorial_elpd} displays the results.

The four models that place a hierarchical prior on $\boldsymbol{\tau}^p_{ij}$ substantially improve on the baseline, the gains ranging between $53.7$ and $60.2$ in cross-validated log predictive density (elpd).  This places them between $6.7$ and $7.9$ standard errors above 0 in paired per-serve differences. The gains from adding hierarchical priors on $\zeta^p_{ij}$ and $\alpha^p_{ij}$ once the $\boldsymbol{\tau}^p_{ij}$ are pooled are much more modest.  Specifically, relative to the $\tau^p$ model, the $\tau^p + \zeta_j + \alpha_i$ model gains $6.5$ in elpd (standard error $4.9$).  We adopt the $\tau^p + \zeta_j + \alpha_i$ model, which has the highest elpd of all the candidate models, while noting that the four models that pool the scales are within noise.

\section{Approximating the Reward Surface} \label{sec:reward_surface}

To approximate $\Pr(R_i=1 \mid \mathbf{x}, \mathbf z_i)$, we leverage model output from \citet{kovalchik2020space} to recursively combine
shot-level win probabilities as follows.  Let $w_t(\mathbf z\mid\mathbf{x})$ denote the probability that the server wins the point
on the $t$th shot of the rally, conditional on the rally
reaching that shot and on game context $\mathbf{x}$
(e.g., deuce/ad court, opponent handedness).
Conditional on a serve landing at $\mathbf z$, the
shot-level model provides an estimate of $w_1(\mathbf z\mid\mathbf{x})$,
the probability of an immediate winning outcome.
With probability $1 - w_1(\mathbf z\mid\mathbf{x})$
the rally continues. In that case, we condition the
trajectory generation model on the serve landing
location and the implied player positions to obtain
the location of the next shot in the rally. At that
new location we evaluate $w_2(\mathbf z\mid\mathbf{x})$,
update the player positions, and proceed recursively.

The overall point-level win probability can then be computed as
\begin{align}
p(R_i = 1 \mid \mathbf{x},\mathbf z)
= \sum_{t=1}^{\infty}
\left(
\prod_{j=1}^{t-1}
(1 - w_j(\mathbf z\mid\mathbf{x}))
\right)
w_t(\mathbf z\mid\mathbf{x}).
\end{align}
We approximate this infinite sum via Monte Carlo
simulation of shot sequences and truncate the recursion
once the continuation probability becomes negligible.   

\section{Additional Figures and Tables} \label{sec:additional_figures}

\vspace{-.1in}

\begin{table}[H]
    \centering
    \footnotesize
    \setlength{\tabcolsep}{3pt}
\begin{tabular}{l|cccc|c || cccc|c}
\toprule
& \multicolumn{5}{c}{1st Serve} & \multicolumn{5}{c}{2nd Serve} \\
\cmidrule(lr){2-6}\cmidrule(lr){7-11}
Player & D.\ Wide & D.\ T & Ad T & Ad Wide & Total & D.\ Wide & D.\ T & Ad T & Ad Wide & Total \\
\midrule
1 & 19 (5) & 24 (1) & 24 (4) & 20 (3) & 87 (13) & 3 (1) & 4 (1) & 5 (0) & 6 (1) & 18 (3) \\
2 & 17 (6) & 16 (5) & 14 (7) & 19 (7) & 66 (25) & 6 (1) & 7 (0) & 5 (0) & 13 (2) & 31 (3) \\
3 & 3 (0) & 7 (1) & 8 (1) & 11 (4) & 29 (6) & 3 (0) & 6 (0) & 2 (0) & 5 (0) & 16 (0) \\
4 & 12 (3) & 12 (6) & 16 (4) & 9 (3) & 49 (16) & 6 (0) & 6 (0) & 5 (0) & 9 (0) & 26 (0) \\
5 & 10 (3) & 12 (2) & 11 (2) & 14 (1) & 47 (8) & 1 (0) & 5 (1) & 2 (0) & 5 (2) & 13 (3) \\
6 & 12 (4) & 7 (0) & 10 (1) & 6 (0) & 35 (5) & 4 (0) & 5 (1) & 1 (0) & 5 (1) & 15 (2) \\
7 & 5 (1) & 11 (3) & 11 (4) & 9 (2) & 36 (10) & 7 (1) & 4 (0) & 6 (0) & 7 (4) & 24 (5) \\
8 & 13 (3) & 15 (5) & 14 (1) & 10 (2) & 52 (11) & 4 (0) & 10 (1) & 6 (0) & 3 (0) & 23 (1) \\
\midrule
Total & 91 (25) & 104 (23) & 108 (24) & 98 (22) & 401 (94) & 34 (3) & 47 (4) & 32 (0) & 53 (10) & 166 (17) \\
\bottomrule
\end{tabular}
     \caption{Serve attempts by player, serve period, and target region. Each entry reports the total number of attempts, including serves censored by net contact; the number of net-censored serves is shown in parentheses. The bottom row reports column totals in the same format.}
    \label{tab:sample_sizes}
\end{table}

\begin{table}[H]
    \centering
    {\footnotesize\setlength{\tabcolsep}{3pt}%
\begin{tabular}{l cc | cc | cc | cc}
\toprule
& \multicolumn{4}{c}{Deuce} & \multicolumn{4}{c}{Ad} \\
\cmidrule(lr){2-5}\cmidrule(lr){6-9}
& \multicolumn{2}{c}{1st Serve} & \multicolumn{2}{c}{2nd Serve} & \multicolumn{2}{c}{1st Serve} & \multicolumn{2}{c}{2nd Serve} \\
\cmidrule(lr){2-3}\cmidrule(lr){4-5}\cmidrule(lr){6-7}\cmidrule(lr){8-9}
Player & Wide & T & Wide & T & Wide & T & Wide & T \\
\midrule
1 & 1.00 & 0.09 & 0.71 & 0.30 & 1.00 & 0.97 & 0.58 & 0.47 \\
2 & 1.00 & 1.00 & 0.93 & 0.82 & 1.00 & 1.00 & 1.00 & 0.80 \\
3 & 1.00 & 0.99 & 0.84 & 0.46 & 1.00 & 1.00 & 0.90 & 0.46 \\
4 & 1.00 & 0.97 & 0.98 & 0.11 & 1.00 & 1.00 & 0.96 & 0.66 \\
5 & 1.00 & 0.96 & 0.95 & 0.17 & 1.00 & 1.00 & 0.81 & 0.81 \\
6 & 1.00 & 0.98 & 1.00 & 0.02 & 1.00 & 1.00 & 0.97 & 0.59 \\
7 & 1.00 & 0.83 & 0.65 & 0.49 & 0.98 & 0.99 & 0.43 & 0.89 \\
8 & 1.00 & 1.00 & 0.99 & 0.55 & 1.00 & 1.00 & 0.96 & 0.89 \\
\midrule
Average & 1.00 & 0.85 & 0.88 & 0.36 & 1.00 & 1.00 & 0.83 & 0.70 \\
\bottomrule
\end{tabular}
}
    \caption{Proportion of the 1000 posterior draws of $\mathcal{M}\bigl(Q_1(\cdot \mid \mathbf{x}, \mathbf{a}_2^{*})\bigr)$ and $\mathcal{M}\bigl(Q_2(\cdot \mid \mathbf{x})\bigr)$ in which a mode existed within each strategy, stratified by player, service court, and serve period.  The bottom row gives the average across players.}
    \label{tab:mode_existence}
\end{table}

\begin{table}[H]
    \centering
    {\footnotesize\setlength{\tabcolsep}{3pt}%
\begin{tabular}{l  ccc | ccc | ccc | ccc}
\toprule
& \multicolumn{6}{c}{Deuce} & \multicolumn{6}{c}{Ad} \\
\cmidrule(lr){2-7}\cmidrule(lr){8-13}
& \multicolumn{3}{c}{1st Serve} & \multicolumn{3}{c}{2nd Serve} & \multicolumn{3}{c}{1st Serve} & \multicolumn{3}{c}{2nd Serve} \\
\cmidrule(lr){2-4}\cmidrule(lr){5-7}\cmidrule(lr){8-10}\cmidrule(lr){11-13}
Player & Wide & T & Diff. & Wide & T & Diff. & Wide & T & Diff. & Wide & T & Diff. \\
\midrule
1 & 0.20 & 0.10 & 0.10 & -0.21 & -0.22 & 0.01 & 0.15 & 0.10 & 0.05 & -0.17 & -0.14 & -0.03 \\
2 & 0.37 & 0.19 & 0.18 & -0.03 & -0.04 & 0.01 & 0.25 & 0.22 & 0.03 & -0.02 & -0.05 & 0.03 \\
3 & 0.38 & 0.26 & 0.12 & 0.04 & 0.03 & 0.01 & 0.25 & 0.20 & 0.05 & 0.00 & -0.03 & 0.03 \\
4 & 0.33 & 0.22 & 0.11 & 0.03 & 0.02 & 0.01 & 0.22 & 0.22 & 0.00 & 0.03 & 0.00 & 0.03 \\
5 & 0.40 & 0.22 & 0.18 & 0.03 & 0.01 & 0.02 & 0.26 & 0.22 & 0.04 & -0.06 & 0.01 & -0.07 \\
6 & 0.43 & 0.24 & 0.19 & 0.15 & 0.06 & 0.09 & 0.25 & 0.20 & 0.05 & 0.03 & -0.01 & 0.04 \\
7 & 0.14 & 0.04 & 0.10 & -0.16 & -0.16 & 0.00 & 0.05 & 0.06 & -0.01 & -0.28 & -0.23 & -0.05 \\
8 & 0.34 & 0.27 & 0.07 & 0.05 & -0.05 & 0.10 & 0.36 & 0.34 & 0.02 & 0.03 & 0.05 & -0.02 \\
\midrule
Average & 0.32 & 0.19 & 0.13 & -0.01 & -0.05 & 0.04 & 0.22 & 0.19 & 0.03 & -0.06 & -0.05 & -0.01 \\
\bottomrule
\end{tabular}
}
    \caption{Value of the objective at each strategy's mode, $Q_i(\mathbf{m}_i^{*}(s \mid \mathbf{x}) \mid \mathbf{x})$ for $s \in \{W, T\}$, and their difference, averaged over draws in which both modes exist, stratified by service court and period. 
    }
    \label{tab:mode_gap}
\end{table}

\begin{figure}[H]
    \centering
    \includegraphics[trim = .3in .15in .2in 0in, clip, width=\textwidth]{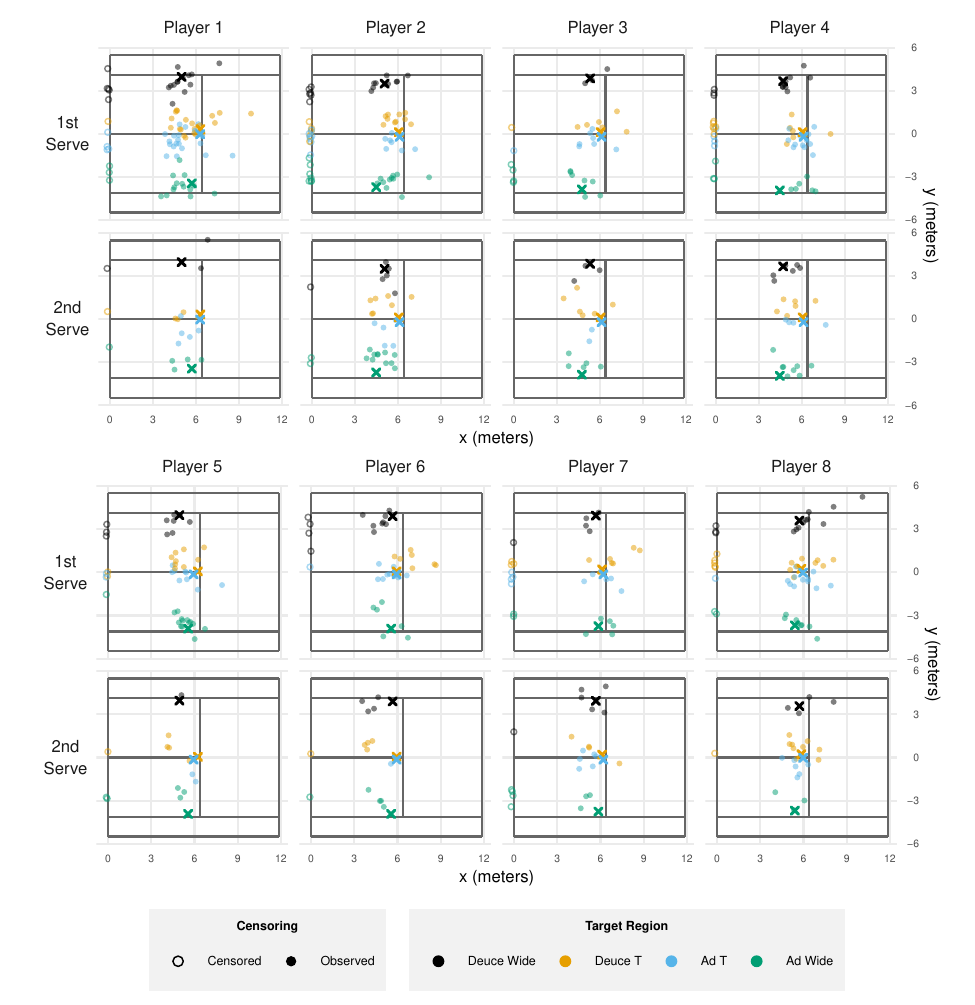}
    \caption{Observed serve data for all eight players in the experiment, stratified by player (columns) and serve period (rows).  X's denote the players' stated target locations, filled circles show observed bounce locations, and unfilled circles represent serves that hit the net (censored), plotted along the net with vertical jitter added for visual clarity.}
    \label{fig:tennis_data_all}
\end{figure}

 \begin{figure}[H]
    \centering
    \includegraphics[trim = .2in 0in .1in 0in, clip, width=1\textwidth]{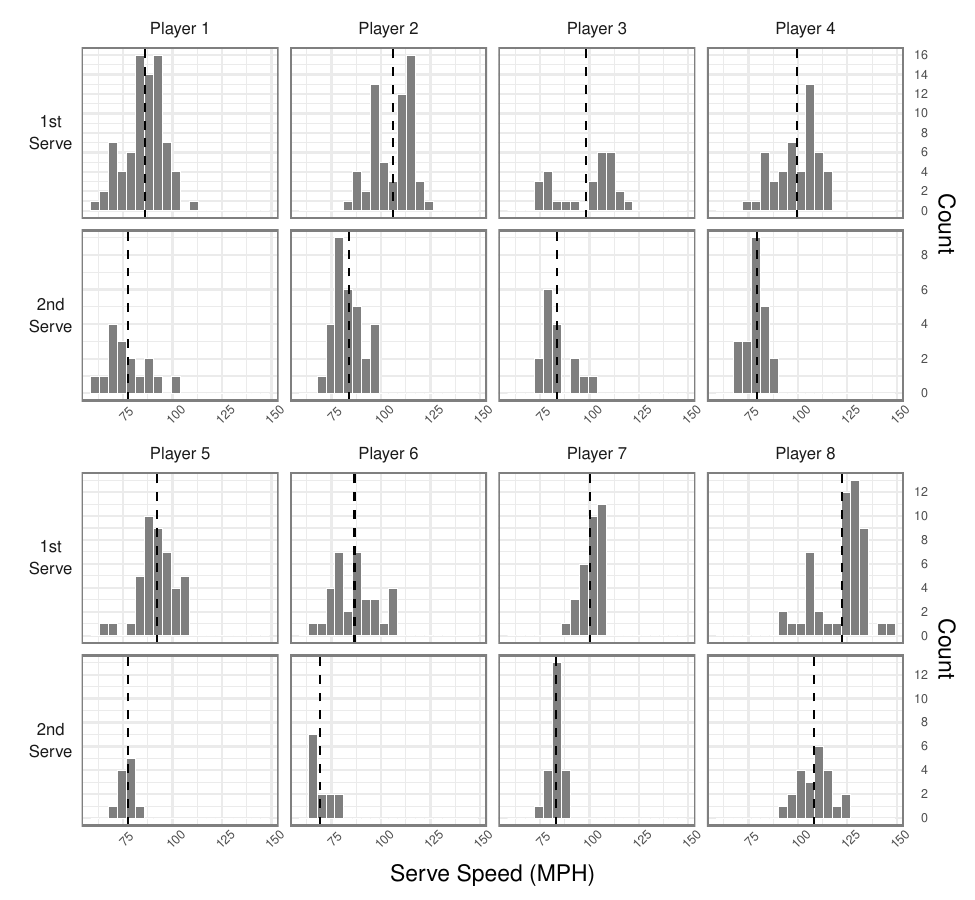}
    \caption{Observed serve speeds for all eight players in the experiment, stratified by player (columns) and serve period (rows).  In each panel, the dashed vertical line indicates the average speed for the serves within that panel.}
    \label{fig:tennis_speeds}
\end{figure}

\begin{figure}[H]
    \centering
    \includegraphics[trim = .0in .0in .0in .0in, clip, width=1\textwidth]{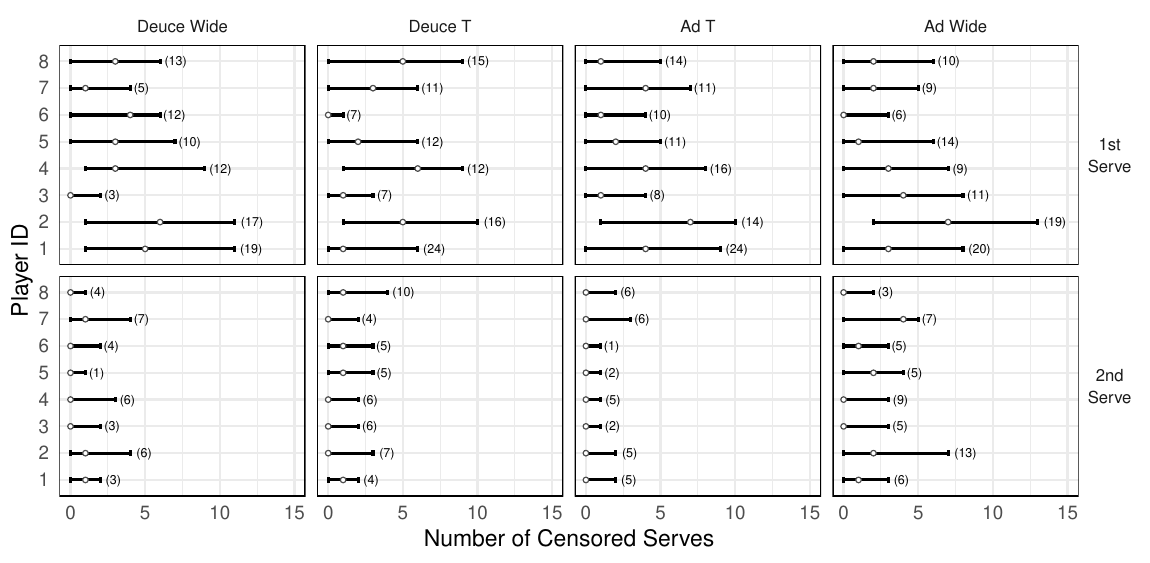}
    \caption{Posterior predictive checks for net serve (censored) counts under the serve execution model. Horizontal bars show 95\% posterior predictive intervals for the number of net serves, and white-filled points indicate observed counts.  The total number of attempts for each player, serve period, region combination are shown in parentheses.}
    \label{fig:ppc_net}
\end{figure}

\begin{figure}[H]
    \centering
    \includegraphics[width=1\textwidth]{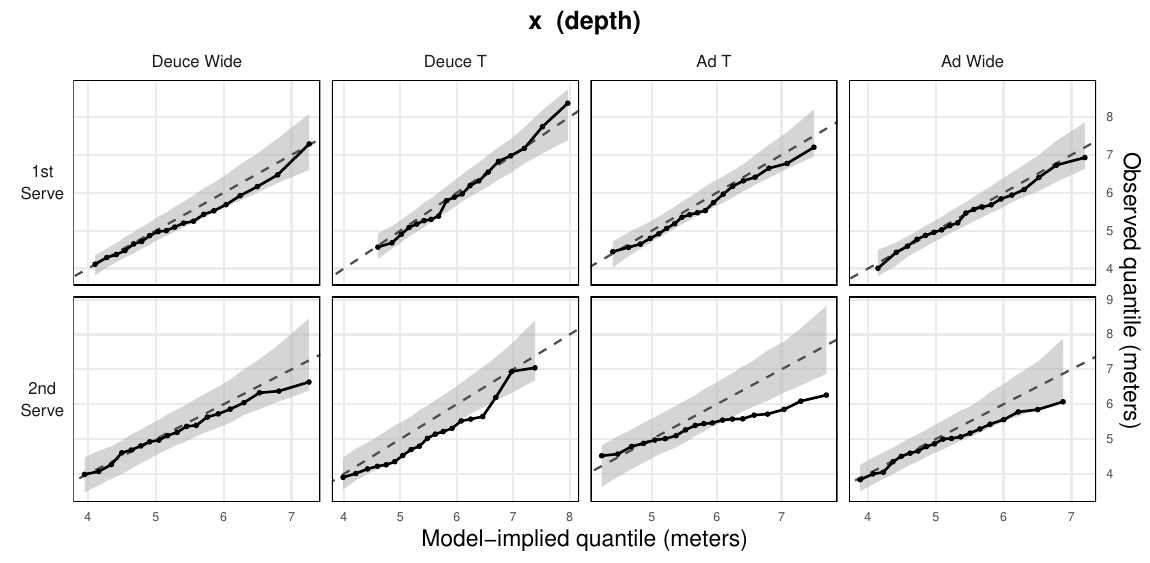}\\[2ex]
    \includegraphics[width=1\textwidth]{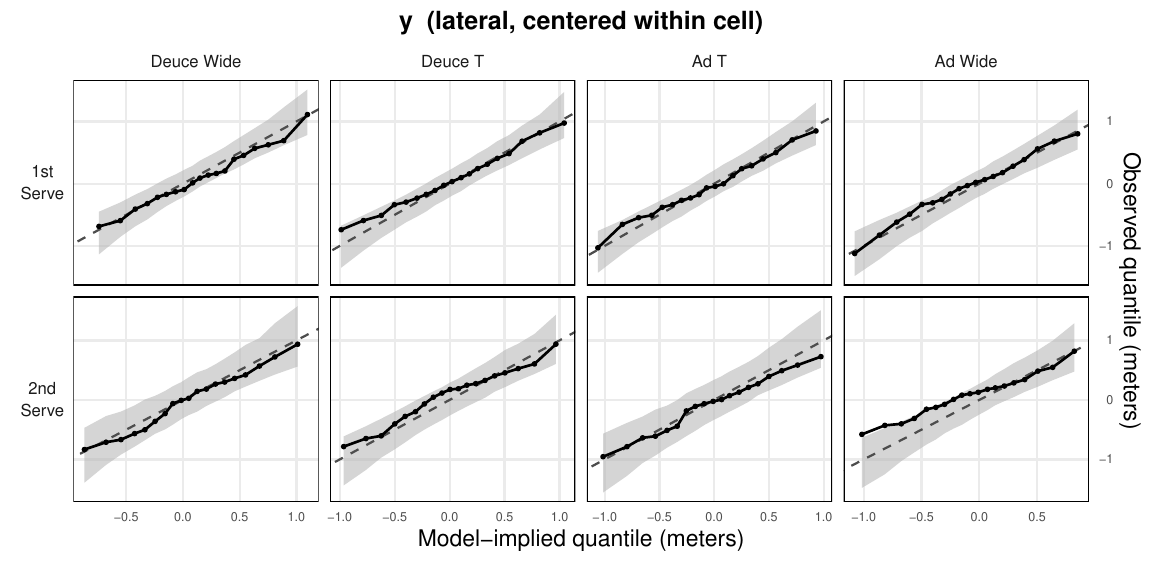}
    \caption{Quantile-quantile posterior predictive checks for serve landing depth ($x$, top) and lateral position ($y$, bottom), stratified by serve period and region. Each panel plots the 5th through 95th percentiles in steps of five. For each of 1{,}000 posterior draws we simulated one serve per observed serve, using that serve's own player, period, and region parameters, so that every replicate has the same cell composition and the same sample size as the data. Within each replicate, we then computed those percentiles for the serves in each of these eight period-region combinations. The horizontal position of each point is the median of the resulting 1{,}000 values, while the gray ribbon is the 95\% interval.  The vertical position is the corresponding observed percentile. Lateral position is shifted by a fixed per-cell constant, subtracted identically from the observed and the simulated serves, so that the four regions share a common scale.}
    \label{fig:ppc_qq}
\end{figure}

\begin{figure}[H]
    \centering
    \includegraphics[width=.95\textwidth]{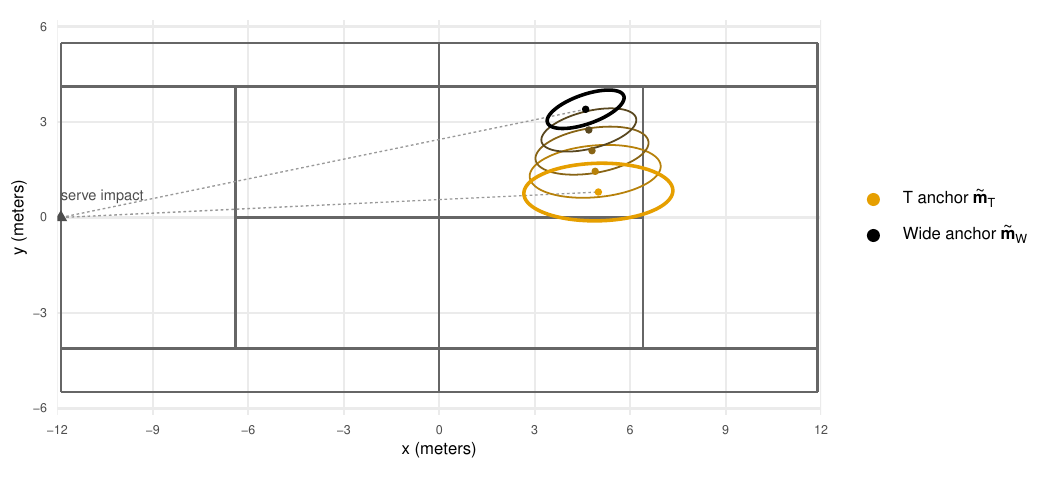}
    \caption{Schematic illustration of the angular interpolation of the serve execution distribution for a Deuce court serve.  The black and gold ellipses are 95\% execution-error contours at the Wide and T anchors ($\widetilde{\mathbf{m}}_W$ and $\widetilde{\mathbf{m}}_T$). The ellipses between these two anchors are interpolated according to the angular interpolation process described in Section~\mainref{sec:interpolation}{4.4}.  Dashed rays mark the angular span between the two anchors.}
    \label{fig:interp_illustration}
\end{figure}

\begin{figure}[H]
    \centering
    \includegraphics[trim = .05in .25in 0in .2in, clip, width=1\textwidth]{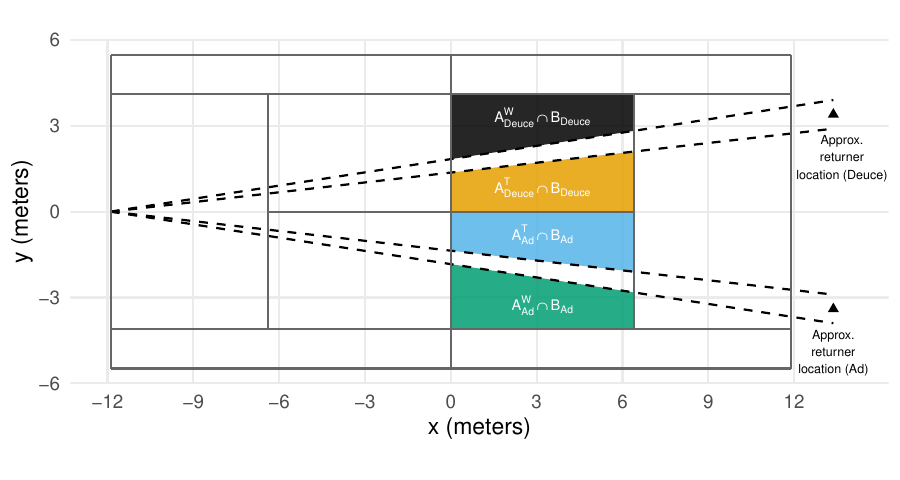}
    \caption{Geometry of the strategy regions on the tennis court. The dashed rays define the Wide, Body, and T aiming regions for each service court. Shaded regions show the intersections of the Wide and T regions with the service boxes.
    }
    \label{fig:action_space}
\end{figure}

\begin{figure}[H]
    \centering
    \includegraphics[trim=.65in .25in .5in .25in clip, width=.9\textwidth]{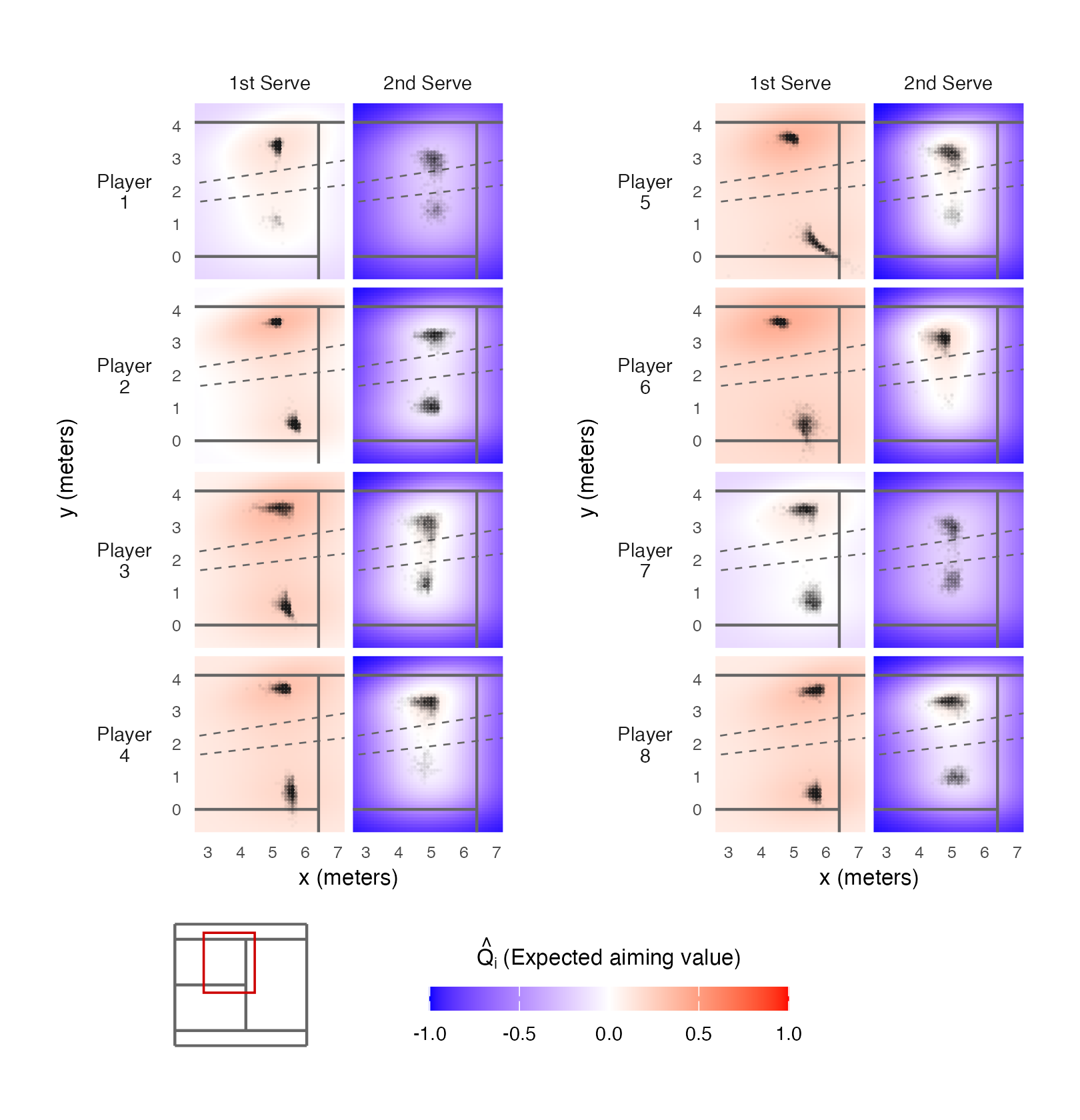}
    \caption{Estimated objective functions $Q_1(\mathbf{a}_1 \mid \mathbf{x}, \mathbf{a}_2^{*})$ and $Q_2(\mathbf{a}_2 \mid \mathbf{x})$ for all eight players on the Deuce court. 
    Black dots denote posterior draws of the corresponding modes, $\mathcal{M}\bigl(Q_1(\cdot \mid \mathbf{x}, \mathbf{a}_2^{*})\bigr)$ and $\mathcal{M}\bigl(Q_2(\cdot \mid \mathbf{x})\bigr)$, for 1000 draws.  Dashed gray lines separate the Wide, Body, and T regions.}
    \label{fig:optimal_aiming_all_du}
\end{figure}

\begin{figure}[H]
    \centering
    \includegraphics[trim=.65in .25in .5in .25in clip, width=.9\textwidth]{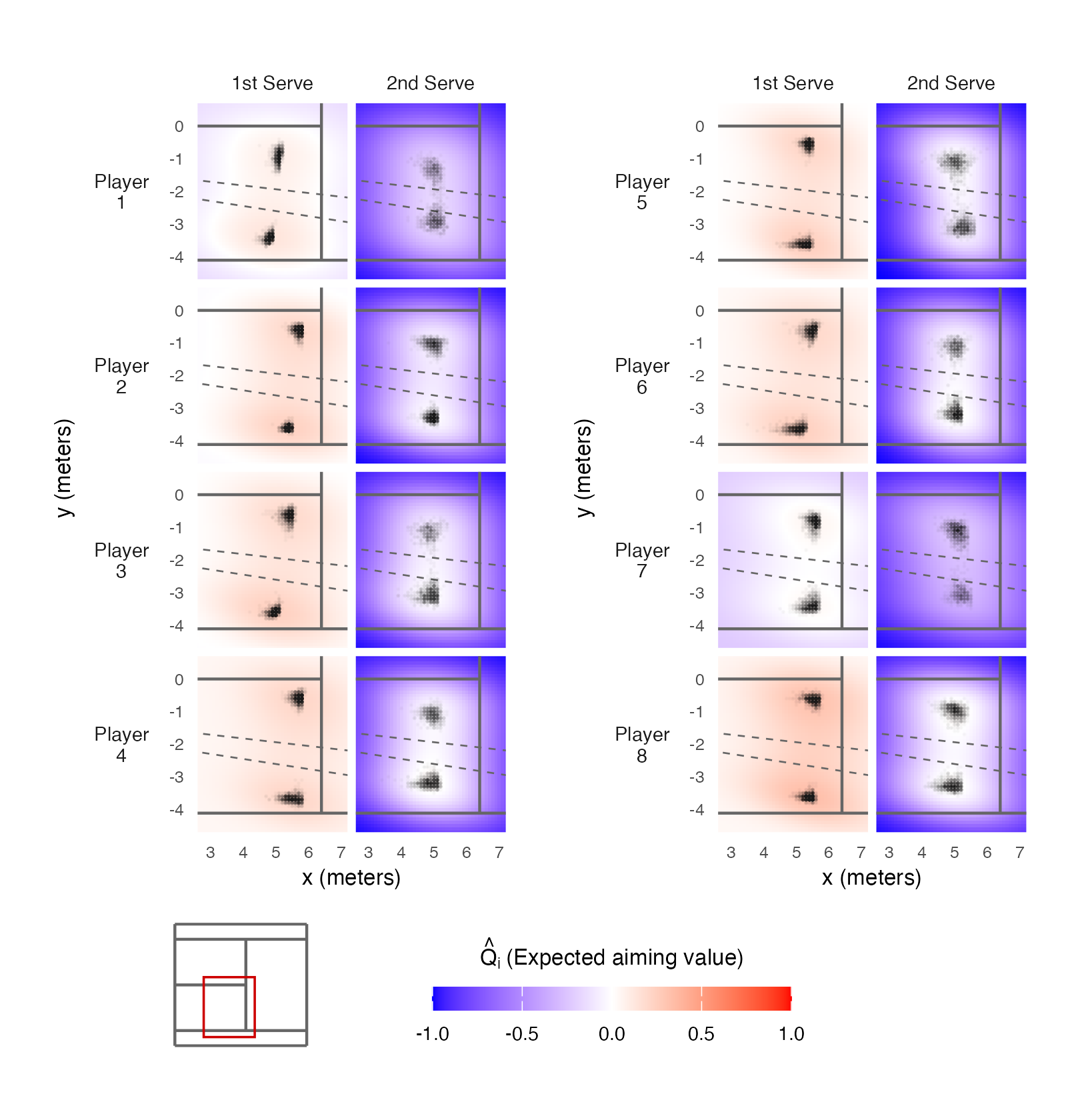}
    \caption{Estimated objective functions $Q_1(\mathbf{a}_1 \mid \mathbf{x}, \mathbf{a}_2^{*})$ and $Q_2(\mathbf{a}_2 \mid \mathbf{x})$ for all eight players on the Ad court. 
    Black dots denote posterior draws of the corresponding modes, $\mathcal{M}\bigl(Q_1(\cdot \mid \mathbf{x}, \mathbf{a}_2^{*})\bigr)$ and $\mathcal{M}\bigl(Q_2(\cdot \mid \mathbf{x})\bigr)$, for 1000 draws.  Dashed gray lines separate the Wide, Body, and T regions.}
    \label{fig:optimal_aiming_all_ad}
\end{figure}

\begin{figure}[H]
    \centering
    \includegraphics[width=0.9\textwidth]{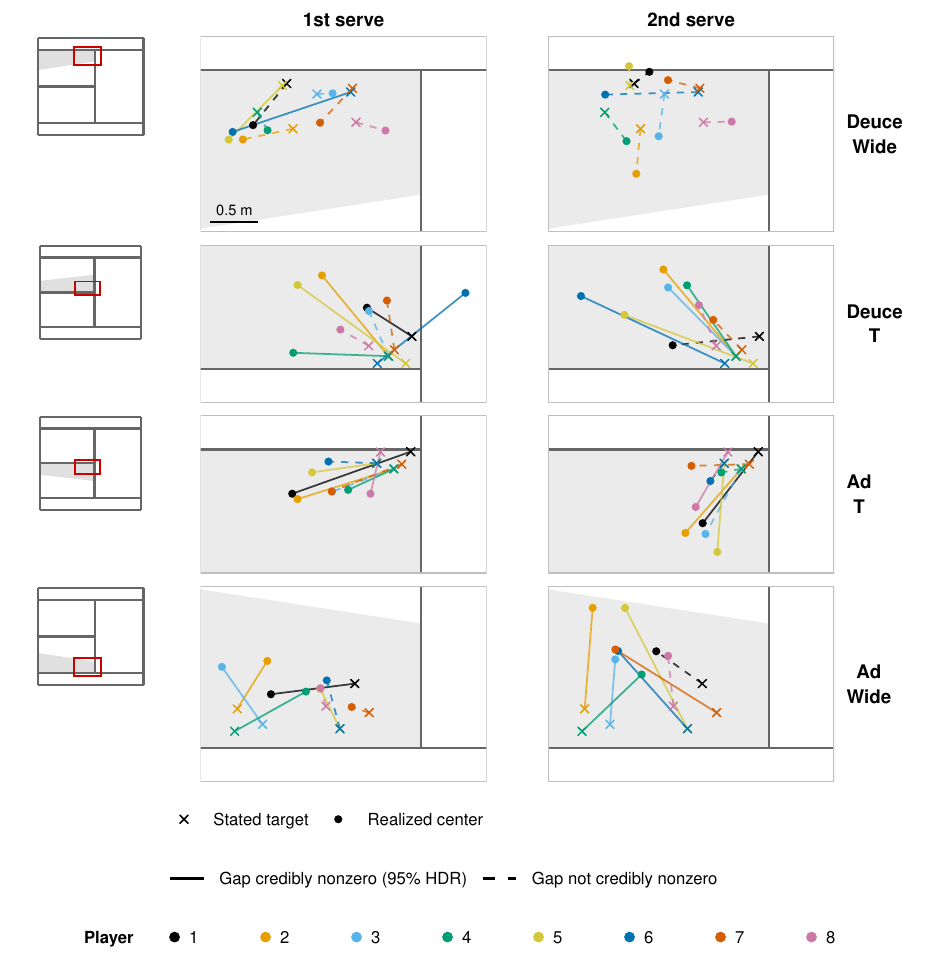}
    \caption{Stated--realized gap for all eight players in the four target regions, displayed as in Figure~\mainref{fig:bias_court_deuce_stated_realized}{9} of the main text. }
    \label{fig:bias_court_stated_realized}
\end{figure}

\begin{figure}[H]
    \centering
    \includegraphics[width=0.78\textwidth]{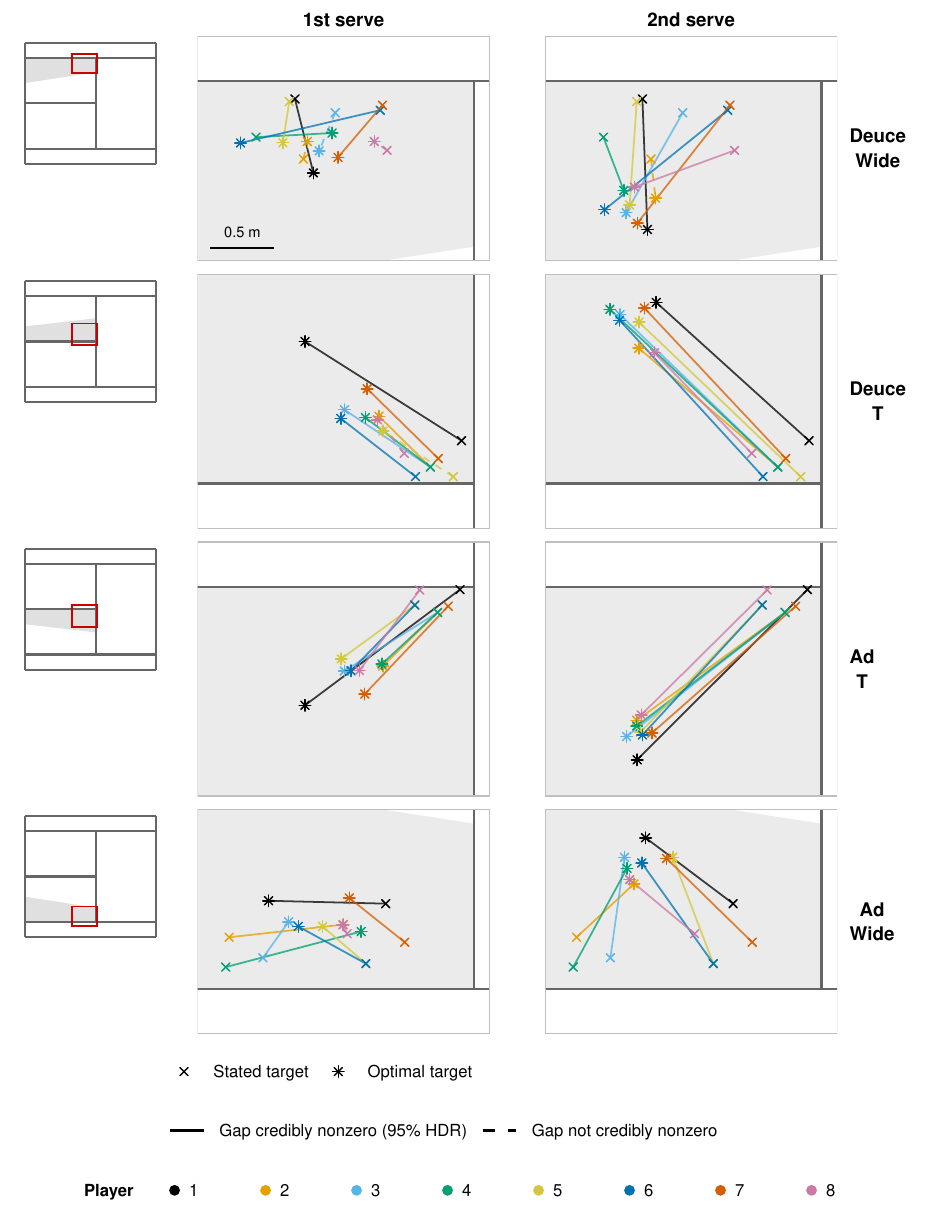}
    \caption{Stated--optimal gap for all eight players in the four target regions, displayed as in Figure~\mainref{fig:bias_court_deuce_stated_optimal}{10} of the main text.  }
    \label{fig:bias_court_stated_optimal}
\end{figure}

\begin{figure}[H]
    \centering
    \includegraphics[width=0.9\textwidth]{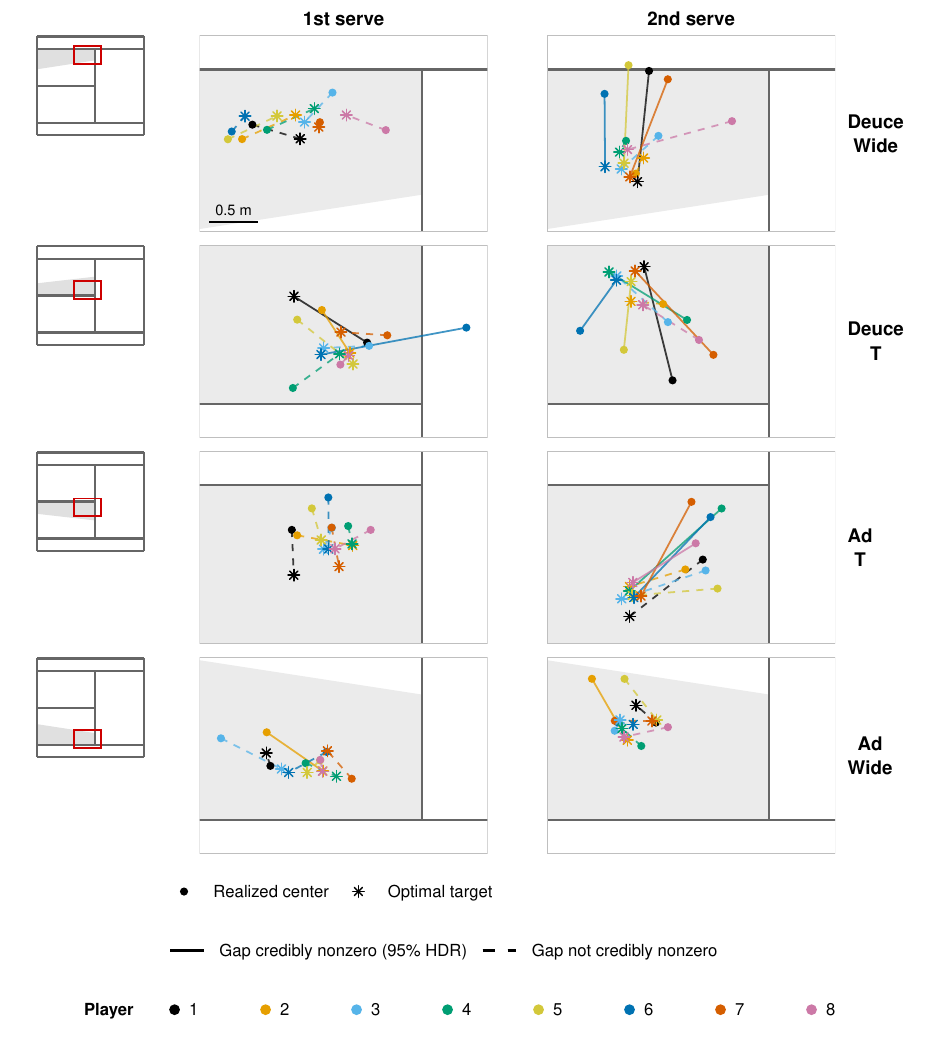}
    \caption{Realized--optimal gap for all eight players in the four target regions, displayed as in Figure~\mainref{fig:bias_court_deuce_realized_optimal}{11} of the main text.}
    \label{fig:bias_court_realized_optimal}
\end{figure}

\end{document}